\documentclass[journal,comsoc]{IEEEtran}
\usepackage{times}

\usepackage{physics}
\usepackage{hyperref}
\usepackage{multirow}
\usepackage{graphicx}
\usepackage[font=footnotesize]{caption}
\usepackage{amsmath}
\usepackage{amsfonts}
\usepackage{amssymb}
\usepackage{amsthm}
\usepackage{tabularx}
\usepackage{mathrsfs} 
\usepackage{cite}

\usepackage{subfig}

\usepackage{soul}

\usepackage{stfloats}
\usepackage{float}
\usepackage{url}
\usepackage{hyperref}

\newtheorem{remark}{Remark}

\usepackage[nolist,nohyperlinks]{acronym}
\usepackage[numbers,sort&compress]{natbib}

\usepackage{bbm}

\usepackage{array}
\usepackage{booktabs}
\usepackage{colortbl}
\usepackage{mdwmath}
\usepackage{mdwtab}
\usepackage{eqparbox}
\usepackage{cellspace}
\usepackage{makecell}

\DeclareGraphicsExtensions{.png,.eps,.ps,.pdf}

\usepackage{tikz}
\usetikzlibrary{arrows,    
                shapes, 
                positioning, 
                fit,    
                backgrounds, 
                mindmap,
                petri,
                intersections, 
                external %
                }
\definecolor{emerald}{RGB}{69,155,61}
\definecolor{gold}{RGB}{244,216,51}
\definecolor{pink}{RGB}{235,44,206}
\usepackage{verbatim}

\begin{acronym}
\acro{AI}{artificial intelligence}
\acro{CE}{channel estimation}
\acro{AMC}{adaptive modulation and coding}
\acro{AWGN}{additive white Gaussian noise}
\acro{BLER}{block error rate}
\acro{CNN}{convolutional neural network}
\acro{RNN}{recurrent neural network}
\acro{SNR}{signal-to-noise ratio}
\acrodefplural{SNR}[SNRs]{signal-to-noise ratios}
\acro{RV}{random variable}
\acrodefplural{RV}[RVs]{random variables}
\acro{NLOS}{non line-of-sight}
\acro{LOS}{line-of-sight}
\acrodefplural{LoS}{Line-of-sight}
\acro{EH}{energy harvesting}
\acro{PDF}{probability density function}
\acro{CDF}{cumulative distribution function}
\acrodefplural{CDF}[CDFs]{cumulative distribution functions}
\acro{IoT}{internet of things}
\acro{PB}{power beacon}
\acrodefplural{PB}[PBs]{power beacons}
\acro{WPC}{wireless powered communications}
\acro{WPT}{wireless power transmission}
\acro{RFID}{radio frequency identification}
\acro{MC}{Monte Carlo}
\acro{PLS}{physical layer security}
\acro{SWIPT}{simultaneous wireless and information power transfer}
\acro{MIMO}{multiple-input multiple-output}
\acro{OFDM}{orthogonal frequency division multiplexing}
\acro{CQI}{channel quality indicator}
\acro{PMI}{precoding matrix indicator}
\acro{RI}{rank indicator}
\acro{LA}{link adaptation}
\acro{CSI}{channel state information}
\acro{TDD}{time division duplexing}
\acro{FDD}{frequency division duplexing}
\acro{FLOPs}{floating point operations}
\acro{RB}{resource block}
\acrodefplural{RB}[RBs]{resource blocks}
\acro{RE}{resource element}
\acrodefplural{RE}[REs]{resource elemets}
\acro{BS}{base station}
\acrodefplural{BS}[BSs]{base stations}
\acro{UE}{user equipment}
\acrodefplural{UE}[UEs]{user equipments}
\acro{LEO}{low earth orbit}
\acro{TDL}{tapped delay line}
\acro{NMSE}{normalized mean squared error}
\acro{MSE}{mean squared error}
\acro{MCS}{modulation and coding scheme}
\acro{SRS}{sounding reference signal}
\acro{DNN}{deep neural network}
\acrodefplural{DNN}[DNNs]{deep neural networks}
\acro{RNN}{recurrent neural network}
\acro{OLLA}{outer loop link adaptation}
\acrodefplural{RNN}[RNNs]{recurrent neural networks}
\acro{LSTM}{long short-term memory}
\acro{GRU}{gated recurrent unit}
\acrodefplural{GRU}[GRUs]{gated recurrent units}
\acro{LS}{least square}
\acro{SINR}{signal to interference plus noise ratio}
\acro{EESM}{exponential effective SINR mapping}
\end{acronym}

\tikzstyle{int}=[draw, fill=cyan!20, minimum size=2em]
\tikzstyle{int_blue}=[draw, fill=black!20, minimum size=2em]
\tikzstyle{int_green}=[draw, fill=green!20, minimum size=2em]
\tikzstyle{int_red}=[draw, fill=red!20, minimum size=2em]
\tikzstyle{init} = [pin edge={to-,thin,black}]

\begin{document}
\title{CSI Prediction Frameworks for Enhanced 5G Link Adaptation: Performance–Complexity Trade-offs}


\author{Francisco D\'iaz-Ruiz, Francisco J. Mart\'in-Vega, Jos\'e A. Cort\'es, Gerardo G\'omez and Mari Carmen Aguayo
\thanks{Manuscript received April xx, 2025; revised XXX. %
This work has been supported by Grant PID2022-137522OB-I00 funded by MCIN/AEI/10.13039/501100011033 and by FEDER {A way of making Europe}, Keysight Technologies, Junta de Andaluc\'ia, and University of M\'alaga (UMA).}
\thanks{The authors are with the Communications and Signal Processing Lab, Telecommunication Research Institute (TELMA), Universidad de M\'alaga, E.T.S. Ingenier\'ia de Telecomunicaci\'on, Bulevar Louis Pasteur 35, 29010 M\'alaga (Spain). (e-mail: \{fdiaz, fjmv, jaca, ggomez, aguayo\}@ic.uma.es)}
}

\markboth{IEEE TRANSACTIONS ON VEHICULAR TECHNOLOGY, VOL. XXX, NO. XXX, MAY 2025}%
{Shell \MakeLowercase{\textit{et al.}}: Bare Demo of IEEEtran.cls for IEEE Communications Society Journals}

\maketitle
\begin{abstract}
Accurate and timely \ac{CSI} is fundamental for efficient link adaptation. However, challenges such as channel aging, user mobility, and feedback delays significantly impact the performance of \ac{AMC}. This paper proposes and evaluates two \ac{CSI} prediction frameworks applicable to both \ac{TDD} and \ac{FDD} systems. The proposed methods operate in the effective \ac{SINR} domain to reduce complexity while preserving predictive accuracy. A comparative analysis is conducted between a classical Wiener filter and state-of-the-art deep learning frameworks based on \acp{GRU}, \ac{LSTM} networks, and a delayed \ac{DNN}. The evaluation considers the accuracy of the prediction in terms of \ac{MSE}, the performance of the system, and the complexity of the implementation regarding \ac{FLOPs}. Furthermore, we investigate the generalizability of both approaches under various propagation conditions. The simulation results show that the Wiener filter performs close to \ac{GRU} in terms of \ac{MSE} and throughput with lower computational complexity, provided that the second-order statistics of the channel are available. However, the \ac{GRU} model exhibits enhanced generalization across different channel scenarios. These findings suggest that while learning-based solutions are well-suited for \ac{TDD} systems where the \ac{BS} handles the computation, the lower complexity of classical methods makes them a preferable choice for \ac{FDD} setups, where prediction occurs at the power-constrained \ac{UE}.
\end{abstract}


\acresetall

\section{Introduction}
\label{sec:Intro}

\subsection{Motivation and Scope}

\IEEEPARstart{L}{ink adaptation} (LA) is a fundamental mechanism in modern wireless communication systems, enabling the dynamic adjustment of transmission parameters based on accurate and timely \ac{CSI}~\cite{Vega2021, Goldsmith2005}. By adapting the \ac{MCS} to instantaneous channel conditions, LA aims to maximize spectral efficiency while maintaining reliable communication within the operating constraints of the system~\cite{Yue2020}.

In \ac{TDD} systems, downlink \ac{CSI} can be inferred from uplink measurements due to the reciprocity of the wireless channel. In contrast, \ac{FDD} systems lack such reciprocity, requiring the \ac{UE} to estimate the downlink \ac{CSI} and report it to the \ac{BS}, thus introducing additional signaling overhead. According to the 3GPP specification~\cite{38.214}, the \ac{CSI} report comprises three key indicators: the \ac{PMI}, \ac{RI}, and the \ac{CQI}. Specifically, the \ac{PMI} identifies the optimal precoding matrix from a predefined codebook, the \ac{RI} indicates the preferred \ac{MIMO} transmission rank, and the \ac{CQI} recommends the most suitable \ac{MCS} based on the channel conditions.

A major challenge in \acsu{LA} based on \ac{AMC} is channel aging, which results from the temporal discrepancy between the estimation of \ac{CSI} and its subsequent use for data transmission in future slots~\cite{Papazafeiropoulos2017}. This mismatch is typically caused by user mobility and processing delays, leading to degraded system performance, suboptimal \ac{MCS} selection, and increased error rates~\cite{Truong2013}. Although transmitting reference signals more frequently mitigates channel aging, this solution incurs a significant overhead that reduces the overall system throughput.

To mitigate the effects of channel aging without additional signaling overhead, predictive \ac{CSI} frameworks have emerged as promising solutions. These approaches leverage temporal and spatial correlations in the wireless channel to predict future channel states, enabling more accurate and timely LA decisions~\cite{Jiang2017}. Predictive \ac{CSI} has demonstrated significant potential to improve link reliability and spectral efficiency, particularly in highly dynamic environments.

Classical channel prediction algorithms, such as Wiener filters and Kalman-based methods, have been widely explored due to their simplicity, interpretability, and low computational requirements. These techniques, while suboptimal for predicting either non-Gaussian or non-linear processes, may still yield an adequate performance for practical purposes and thus remain appealing for real-time or resource-constrained implementations. For example, the Kalman filter, which can be regarded as an adaptive linear predictor, has been applied for channel tracking in fading environments~\cite{Liu1995}, and more recent studies have compared Kalman-based predictors against machine learning approaches in Rayleigh fading and massive MIMO systems~\cite{Jiang2019, Kim2021}.

In parallel, traditional 5G systems often employ the \ac{OLLA} mechanism as a reactive link adaptation strategy. \ac{OLLA} dynamically adjusts the selected \ac{MCS} based on the observed \ac{BLER} of previously transmitted blocks, compensating for inaccuracies in the instantaneous \ac{CSI} estimation. Although this approach enhances robustness by maintaining the target error rate, it reacts to past transmission outcomes rather than anticipating future channel variations. Consequently, under fast-varying or highly mobile conditions, \ac{OLLA} may become suboptimal compared to predictive approaches that estimate future channel states~\cite{blanquez2016eolla}.

Recent advances in artificial intelligence (AI) and machine learning (ML) are reshaping the design of link adaptation and channel prediction in 5G and beyond. AI-based models are capable of learning complex, non-linear channel dynamics directly from data, enabling more adaptive and robust link management strategies. Deep learning architectures, such as \acp{CNN} and \acp{RNN}, have shown promising results in tasks such as feedback reduction and mobility-aware channel prediction~\cite{Ye2018, Liao2019}. In future wireless systems, AI is expected to be a key enabler for intelligent radio resource management, end-to-end network optimization, and fully autonomous link adaptation under uncertain or dynamic conditions~\cite{Saad2020}.

The increasing prevalence of AI-based solutions in the literature may stem from their non-linear nature, which allows them to outperform classical methods that are inherently suboptimal for predicting either non-Gaussian or non-linear processes. However, even in such cases, linear techniques may still yield competitive performance for the considered application. This gap highlights the need for a systematic evaluation of both paradigms in terms of prediction accuracy, computational complexity, and generalization capabilities, to better inform practical deployment decisions in diverse network environments.

\subsection{Related Work}

The integration of AI into wireless communications has led to significant advancements in \ac{LA} strategies. In particular, \acp{DNN} have been extensively employed to address various challenges in \ac{LA}, such as optimizing the interval of \ac{CSI} feedback reports to reduce signaling overhead~\cite{2022Hong}.

\acp{RNN}, especially \ac{LSTM} architectures, have demonstrated strong performance in time-series prediction tasks and have been widely adopted for \ac{CSI} prediction~\cite{li2019ea}. In~\cite{Kadambar2023Deep}, for instance, the authors focus on optimizing the trade-off between CSI accuracy and feedback overhead by introducing \ac{LSTM}-based autoencoders for joint compression and prediction.

Several deep learning-based predictive \ac{CSI} techniques have been proposed to mitigate the adverse effects of channel aging in fast-fading environments. Most of these approaches operate directly on the channel matrix, which significantly increases prediction complexity. For instance, in~\cite{2023Gao}, the authors address the challenges of channel aging by proposing a prediction method based on Temporal Convolutional Networks (TCNs) to forecast future channel conditions and select the \ac{CQI}. However, working with full channel matrices to infer \ac{CQI} levels results in a higher computational cost. Similarly,~\cite{2021Yuan} introduces a deep learning framework that predicts future channel matrices and subsequently derives the corresponding \ac{PMI}/\ac{RI} values for link adaptation.

Despite these advancements, a common limitation of existing deep learning approaches is their inherent computational complexity, often arising from operations conducted in the time-frequency domain or through intricate network architectures. Advanced models leveraging Transformers~\cite{Jian2022}, TCNs~\cite{2023Gao}, or emerging concepts like Kolmogorov-Arnold Networks (KANs)~\cite{Vaca2024} often achieve impressive prediction accuracy but at the cost of significant computational resources, memory consumption, and increased inference latency, which might prevent their application to real-time deployments. Furthermore, a gap in the existing literature is the lack of direct comparisons between these sophisticated DNN-based solutions and classical prediction techniques. While some works, such as~\cite{Kadambar2023Smart}, aim to reduce complexity by operating in alternative domains (e.g., mutual information), they often fail to benchmark their performance against conventional predictors that may offer competitive accuracy with substantially lower overhead. Moreover, many studies focus on specific system aspects, such as exploiting partial reciprocity in \ac{FDD} systems~\cite{Qin2022} or integrating joint \ac{CSI} compression and prediction~\cite{Kadambar2023Deep}, which, while valuable, represent specific optimizations rather than a general guide for framework selection.


\subsection{Main Contributions}
In this work, we address the aforementioned gaps by proposing a unified \ac{CSI} prediction framework that operates in the effective \ac{SINR} domain. This approach significantly reduces computational complexity compared to direct \ac{CSI} matrix prediction and is fully compatible with standard \ac{EESM}-based \ac{LA}, enabling low-complexity deployment across diverse scenarios (including terrestrial, vehicular, and non-terrestrial communications). We present a systematic comparison between deep learning frameworks and the classical Wiener filter, focusing on generalization capabilities and key performance metrics. The main contributions of this paper are summarized as follows:

\begin{itemize}
    \item We propose a computationally efficient \ac{CSI} prediction framework that operates in the effective \ac{SINR} domain, making it directly compatible with standard 5G link adaptation procedures like \ac{EESM}-based \ac{CQI} selection. The framework is applicable to a wide range of deployment scenarios (including terrestrial, vehicular, and non-terrestrial communications) where \ac{CSI} feedback latency or overhead is a limiting factor.

    \item Two prediction approaches, which model complex channel dynamics, are developed and compared: (i) a classical Wiener filter, and (ii) deep learning-based predictors using state-of-the-art models, including \ac{GRU}, \ac{LSTM}, and delayed DNN. 

    \item A detailed investigation of key design aspects is presented, including input sampling, the influence of the Doppler-\ac{CSI} reporting product (\( f_D \times T_{\mathrm{CSI}} \)), and the choice of prediction target (full \ac{CQI} vector vs. best \ac{CQI}).

    \item A comprehensive performance assessment is conducted under diverse mobility and channel conditions. The results reveal a clear trade-off: while the Wiener filter provides a competitive low-complexity baseline, the \ac{GRU}-based framework offers enhanced prediction accuracy and superior generalization capabilities, particularly in dynamic and mismatched channel environments.
\end{itemize}

The rest of this paper is organized as follows: Section~II explains the system model and the principle of \ac{CQI} selection. The proposed prediction frameworks are introduced in Section~III, while Section~IV presents the numerical and simulation results. Finally, Section~V provides the conclusions drawn from this research.


\section{System Model}
We consider a \ac{MIMO}-\ac{OFDM} system with $N_{\text{Tx}}$ transmit antennas and $N_{\text{Rx}}$ receive antennas. The system supports both \ac{TDD} and \ac{FDD} CSI acquisition modes in 5G New Radio (NR) systems. The subcarrier spacing is set to $15$~kHz, corresponding to numerology~$0$ in the 5G NR standard.

The time-frequency resource grid is organized into slots of 1~ms duration, each composed of 14 consecutive \ac{OFDM} symbols. The fundamental unit for resource allocation is the \ac{RE}, defined by a single subcarrier in the frequency domain and a single \ac{OFDM} symbol in the time domain. A \ac{RB} consists of 12 contiguous subcarriers. The total system bandwidth is determined by the number of \acp{RB}, denoted as $N_{\text{RB}}$.

The system model consists of the following key components and functionalities:

\begin{itemize}
    \item \textbf{BS:} The \ac{BS} is responsible for managing data and control signal transmissions. In FDD mode, it dynamically adapts link parameters based on feedback from the \ac{UE}. In TDD mode, it leverages channel reciprocity to perform \ac{CE}, acquire \ac{CSI}, and adjust link parameters accordingly.

    \item \textbf{\ac{UE}:} In FDD mode, the \ac{UE} receives the \ac{CSI} reference signal (CSI-RS), performs downlink \ac{CE}, and generates \ac{CSI} feedback for the \ac{BS}. In TDD mode, its primary role is to transmit  \ac{SRS}, which allow the \ac{BS} to estimate the downlink channel via reciprocity.

    \item \textbf{Channel:} In \ac{TDD} systems, channel reciprocity is assumed between the uplink and downlink. In \ac{FDD} systems, the downlink channel is explicitly modeled, while the uplink is considered ideal, i.e., error-free transmissions of CSI reports from the UE to the BS, as our focus is on downlink link adaptation. We consider the 3GPP \ac{TDL} channel models that are recommended for link-level simulations in~\cite{38.901}.
\end{itemize}


\subsection{CSI acquisition framework in 5G}

In this section, we describe the \ac{MCS} selection process at the \ac{BS} in a 5G system and highlight the operational differences between the \ac{TDD} and \ac{FDD} duplexing modes.

\begin{figure}[htbp]
    \centering
    \includegraphics[width=0.95\linewidth]{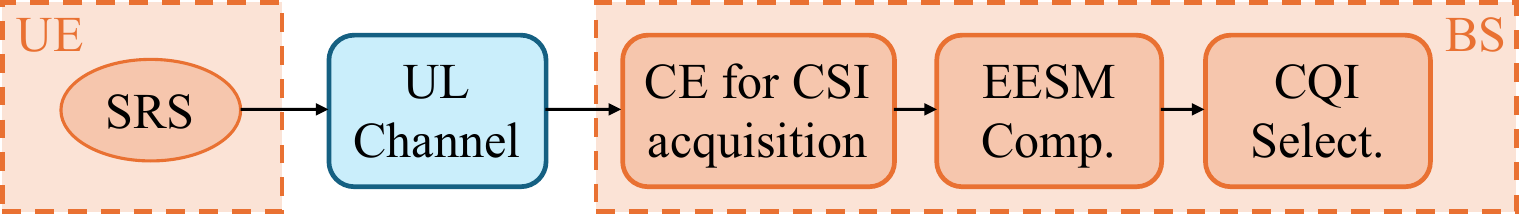}
    \caption{CSI acquisition process in a TDD system.}
    \label{fig:TDDSystem}
\end{figure}

In TDD systems (see Fig.~\ref{fig:TDDSystem}), channel reciprocity between the uplink and downlink is assumed, enabling the \ac{BS} to estimate the downlink channel using \acp{SRS} transmitted by the \ac{UE}. The process begins with the \ac{UE} transmitting \ac{SRS} sequences, which traverse the wireless channel and are received by the \ac{BS}. Leveraging this reciprocity, the \ac{BS} performs \ac{CE} for the downlink directly from the uplink measurements.

Once the channel is estimated, the \ac{BS} computes the effective channel quality using the \ac{EESM} technique. Based on this value, the most suitable \ac{CQI} index is selected to determine the \ac{MCS} for the subsequent downlink transmission. Since the \ac{BS} directly derives the required channel information, no explicit \ac{CSI} feedback from the \ac{UE} is needed, thus reducing signaling overhead and latency. The \ac{UE} then demaps the received \acp{RE}, applies channel equalization, and decodes the transmitted data. System performance is typically evaluated using throughput metrics, which account for successful data delivery after error correction and retransmissions.

\begin{figure}[htbp]
    \centering
    \includegraphics[width=0.9\linewidth]{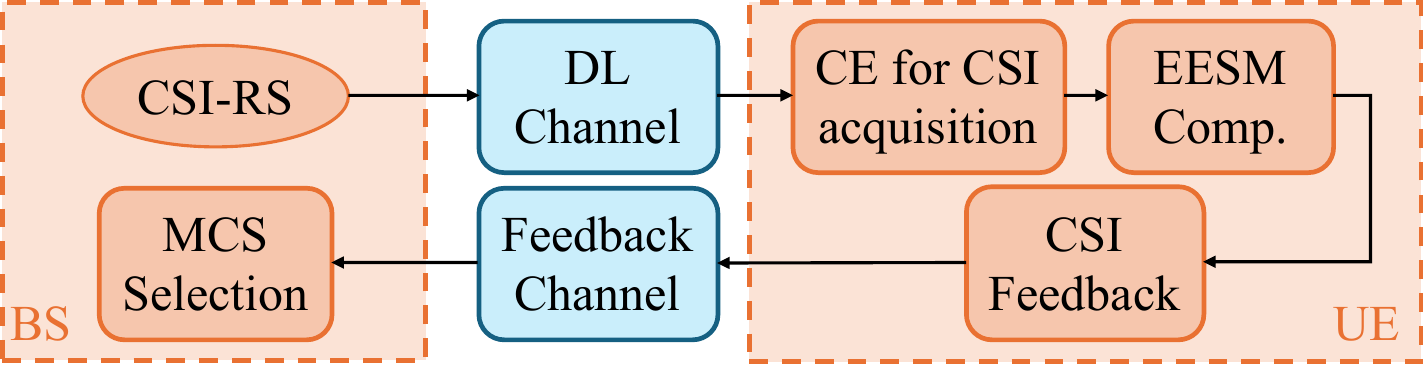}
    \caption{CSI acquisition process in an FDD system.}
    \label{fig:FDDSystem}
\end{figure}

In contrast, \ac{FDD} systems (see Fig.~\ref{fig:FDDSystem}) lack inherent channel reciprocity. Consequently, a feedback loop is required to inform the \ac{BS} of the downlink channel conditions. In this setup, the \ac{BS} transmits CSI-RS signals, which the \ac{UE} receives and uses to perform downlink \ac{CE}. The \ac{UE} then applies the \ac{EESM} method to compress the channel information and generates \ac{CSI} feedback indicators, such as the \ac{CQI}, \ac{PMI}, and \ac{RI}. These indicators are transmitted back to the \ac{BS} via a dedicated uplink control channel. Critically, unlike in \ac{TDD} where the \ac{BS} has access to the continuous effective \ac{SINR}, in \ac{FDD} the \ac{BS} only receives this quantized feedback. This distinction in data availability fundamentally influences the design of the prediction framework utilized for each mode.

Upon receiving the \ac{CSI} feedback, the \ac{BS} selects the appropriate \ac{CQI} and configures the downlink \ac{MCS} accordingly. While this approach enables accurate adaptation to channel conditions, it introduces additional signaling overhead and latency. As in the TDD case, the \ac{UE} performs demapping and equalization to recover the transmitted data, and overall performance is assessed using throughput metrics.

\begin{figure}[htbp]
    \centering
    \includegraphics[width=0.9\linewidth]{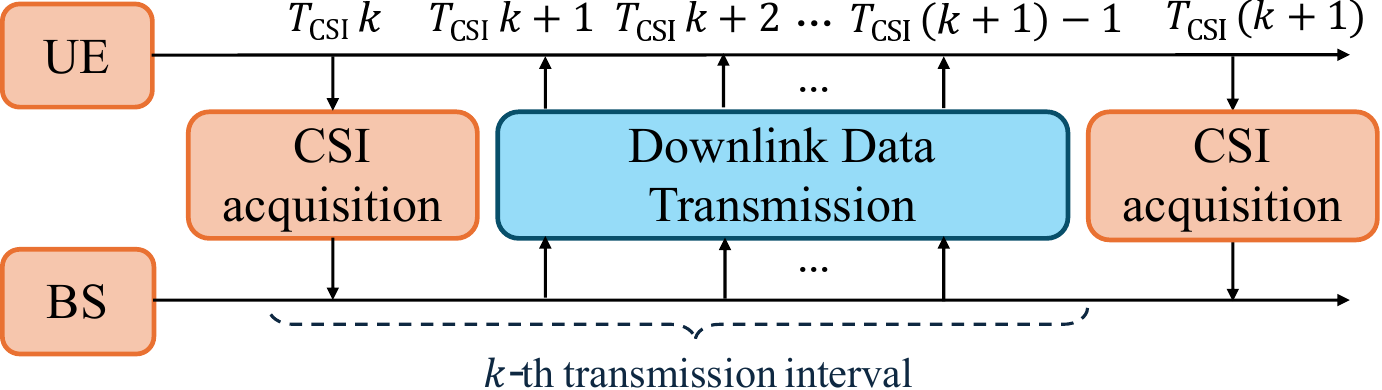}
    \caption{Timeline of the CSI acquisition and downlink data transmission process, illustrating the $k$-th transmission interval.}
    \label{fig:transmissionInterval}
\end{figure}

As discussed in Fig. \ref{fig:transmissionInterval}, the communication is divided into different \textit{transmission intervals} where a \ac{CSI} acquisition stage is performed prior of data transmission. If $T_\mathrm{CSI}$ represents the CSI reporting period (in slot). The data transmission stage at the $k$-th transmission period lasts $T_\mathrm{CSI}-1$ slots, and it uses the \ac{CSI} information obtained from slot $n=T_\mathrm{CSI} k$. Therefore the data transmission at slot, $n=T_\mathrm{CSI} k + \tau$, with $\tau \in [1, T_\mathrm{CSI}-1] \subset \mathbb{N}$, uses outdated CSI by $\tau$ slots, which yields a suboptimal \ac{MCS} selection for those data transmissions. 


During the \ac{CSI} acquisition process, the \ac{SINR} is calculated for each \ac{RB} across the time-frequency grid. However, link adaptation decisions, such as \ac{CQI} selection, require a single representative value that summarizes the overall channel quality at a given time instant. In MIMO systems, signal quality may vary not only across frequency (i.e., \acp{RB}) but also across spatial layers. To consolidate this multidimensional information into a scalar metric, the \ac{EESM} technique is applied jointly across all layers and resource blocks. For each candidate \ac{CQI} level $i \in [1, N_\mathrm{CQI}] \subset \mathbb{N}$, being $N_\mathrm{CQI}=15$ the number of CQIs, the effective SINR at time slot $n$ is computed as \cite{2010Tobias}:

\begin{equation} \label{eq:eesm}
    \gamma^{(i)}_{\mathrm{eff}}(n) = -\beta^{(i)} \ln \left( \frac{1}{N_{\mathrm{L}} N_{\mathrm{RB}}} \sum_{l=1}^{N_{\mathrm{L}}} \sum_{m=1}^{N_{\mathrm{RB}}} \exp\left( -\frac{\gamma_n(l,m)}{\beta^{(i)}} \right) \right),
\end{equation}
where \( \gamma_n(l,m) \) denotes the instantaneous SINR at time slot $n$ on the \( l \)-th spatial layer and \( m \)-th resource block, \( N_{\mathrm{L}} \) is the number of layers, \( N_{\mathrm{RB}} \) is the number of resource blocks, and \( \beta^{(i)} \) is a \ac{CQI}-dependent calibration parameter. The value of \( \beta^{(i)} \) is determined via link-level simulations to ensure that the \ac{EESM} output aligns with the \ac{BLER} performance observed under \ac{AWGN} \cite{Lagen2020}.

Once \( \gamma^{(i)}_{\mathrm{eff}}(n) \) is computed for each \ac{CQI} level at slot $n$, it is mapped to the corresponding \ac{BLER} using precomputed \ac{AWGN} reference curves. This establishes a one-to-one relationship between each candidate \ac{CQI} index and its associated effective SINR value. The final \ac{CQI} index for that slot, \( i^*(n) \), is selected as the highest value whose effective SINR yields a \ac{BLER} below a predefined threshold:

\begin{equation} \label{eq:cqi_selection}
    i^*(n) = \max \left\{ i \,:\, \mathrm{BLER} \left( \gamma^{(i)}_{\mathrm{eff}}(n) \right) \leq \mathrm{BLER}_{\mathrm{target}} \right\}.
\end{equation}

This procedure enables the \ac{BS} to identify the most spectrally efficient \ac{CQI} level—and therefore the corresponding \ac{MCS}—that satisfies the reliability constraint. The selected index \( i^*(n) \) implicitly defines a unique effective SINR, \( \gamma^{(i^*(n))}_{\mathrm{eff}}(n) \). The \ac{EESM}-based method aggregates \ac{SINR} values across the frequency and spatial domains into a single scalar metric, ensuring consistent and robust link adaptation in dynamic wireless environments.

\subsection{Metrics}
In this paper, we evaluate performance using two key metrics: \ac{MSE} and throughput.

\begin{itemize}
    \item \textbf{MSE:} Used to quantify the accuracy of the model's predictions against the ground-truth values. Measures the average of the squared differences between the predicted values and the actual values. The MSE is defined as:
    \begin{equation}
        \text{MSE} = \frac{1}{N} \sum_{i=1}^{N} (y_i - \hat{y}_i)^2,
    \end{equation}
    where \( y_i \) represents the actual values, \( \hat{y}_i \) denotes the predicted values, and \( N \) is the total number of samples.
    \item \textbf{Throughput:} Represents the effective data transmission rate achieved by the system. In this work, throughput refers to the volume of successfully delivered and usable data, excluding retransmissions and protocol overhead. Two forms of throughput are considered:
    \begin{itemize}
        \item \textbf{Conditioned Throughput:} The throughput measured $\tau$ slots after the last CSI acquisition slot, i.e., conditioned on slot $k T_\mathrm{CSI}+\tau$. It captures the temporal evolution of system performance and reflects how prediction accuracy and channel aging affect link adaptation decisions in real time.
        
        \item \textbf{Unconditioned Throughput:} The average throughput computed over the entire simulation duration. It provides a global performance indicator of the prediction strategy's long-term efficiency, integrating the effects of all transmission slots.
    \end{itemize}
\end{itemize}


\section{Proposed \ac{CSI} Prediction Frameworks}
\label{sec:ProposedFrameworks}

The temporal evolution of wireless channels may exhibit correlation over time, enabling inference of future channel states from past observations. To exploit this property, we propose a unified prediction framework that estimates future \ac{CSI} values based on historical data. Instead of operating directly on raw channel coefficients, our framework performs prediction in the effective \ac{SINR} domain ($\gamma_{\mathrm{eff}}$), using the compressed representation obtained through the \ac{EESM} method. This approach yields a compact, standard-compatible, and computationally efficient input well-suited for both classical and learning-based prediction methods.

\begin{figure}[htbp]
    \centering
    \includegraphics[width=0.9\linewidth]{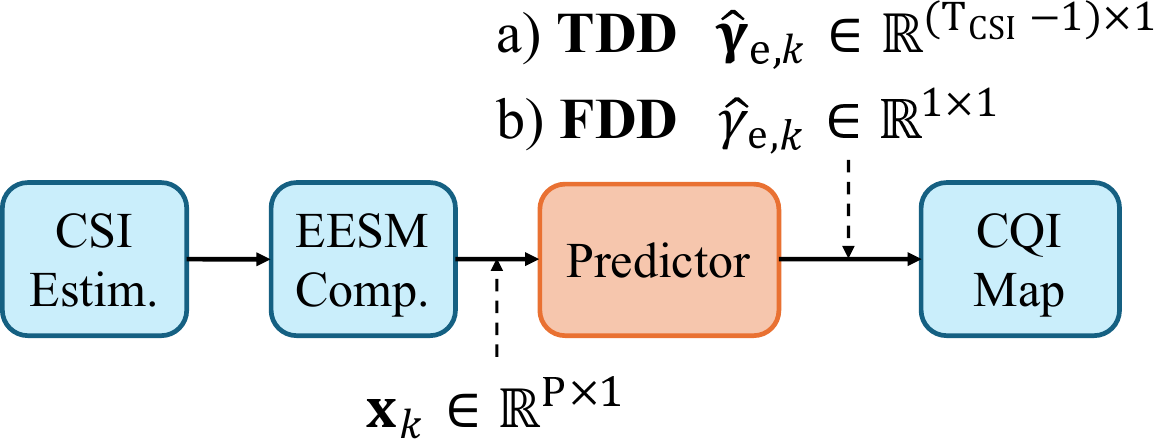}
    \caption{Proposed \ac{CSI} prediction framework.}
    \label{fig:CSIPredictionFramework}
\end{figure}

Fig.~\ref{fig:CSIPredictionFramework} illustrates the overall prediction process. The procedure begins with \ac{CSI} estimation and \ac{EESM} compression, which generates an input vector \( \mathbf{x}_k \) for the predictor at each reporting instant \( k \). To allow a fair comparison between models using the \ac{MSE}, predictions are generated from a standardized version of the effective \ac{SINR}, $\gamma_{\mathrm{e}}(n) = \frac{\gamma_{\mathrm{eff}}(n)-\mathbb{E}\left[\gamma_{\mathrm{eff}}(n) \right]}{\sigma_{\gamma_{\mathrm{eff}}}}$, where \(\sigma_{\gamma_{\mathrm{eff}}}\) is the standard deviation of \(\gamma_{\mathrm{eff}}\). Hence, the vector \( \mathbf{x}_k \) is constructed from a sequence of the $P$ most recent \ac{SINR} samples,
\begin{equation} \label{eq:input_vector}
    \mathbf{x}_k =  
        \left[ \gamma_{\mathrm{e}}(T_\mathrm{CSI} k), \gamma_{\mathrm{e}}(T_\mathrm{CSI}(k-1)), \ldots, \gamma_{\mathrm{e}}(T_\mathrm{CSI}(k-P+1)) \right]^\top.
\end{equation}

The predictor's output depends on the duplexing mode. In \ac{TDD} systems, the framework forecasts a sequence of future effective \ac{SINR} values for the upcoming slots until the next report:
\begin{equation}
\hat{\boldsymbol{\gamma}}_{\mathrm{e}, k} = 
\left[ \hat{\gamma}_{\mathrm{e}}(T_\mathrm{CSI} k+1), \ldots, 
\hat{\gamma}_{\mathrm{e}}(T_\mathrm{CSI} k + T_\mathrm{CSI} - 1)
\right]^\top,
\label{eq:outputSINR}
\end{equation}
which are subsequently mapped to their corresponding \ac{CQI} levels for enhanced link adaptation. In \ac{FDD} systems, the \ac{UE} reports a feedback indicators (\ac{CQI}, \ac{PMI}, and \ac{RI}) intended to guide transmission during the upcoming transmission interval. Consequently, the challenge lies in determining the most appropriate feedback parameters to be used over that future period. Therefore, the goal is to predict a single effective \ac{SINR} value, \( \hat{\gamma}_{\mathrm{e}}(n+\breve{\tau}) \), where the \textit{prediction horizon} \( \breve{\tau} \) is a design parameter that can be optimized to maximize system throughput.

The predictor can operate on two different inputs, depending on the desired modeling granularity. The first approach uses the full vector of effective \ac{SINR} values associated with all available \ac{CQI} levels, providing a multidimensional view of the link quality. This option is referred to as \textbf{by-CQI} and involves performing prediction of the effective \ac{SINR} associated with each CQI, $\hat{\mathbf{\gamma}}^{(i)}_{\mathrm{e},k}$. Alternatively, the predictor can operate on a scalar input, using only the effective \ac{SINR} corresponding to the best-performing \ac{CQI} at each time instant. This latter formulation, which is referred to as \textbf{best-CQI}, is more compact and computationally efficient, highlighting a trade-off between predictive accuracy and computational cost.

\subsection{Wiener Filter Prediction}
\label{subsec:WienerFilter}

As a classical baseline, we implement a discrete-time Wiener filter for linear \ac{CSI} prediction~\cite{kay1993fundamentals}. Let \( \gamma_{\mathrm{e}}(n) \) denote the effective SINR observed in the slot \( n \), assuming that the process is stationary by a wide-sense and has zero mean. The goal is to predict a set of future values \( \gamma_{\mathrm{e}}(n + \tau) \) for \( \tau \in \{1, 2, \ldots, T_{\mathrm{CSI}}-1\} \).

A bank of \( T_{\mathrm{CSI}} - 1 \) linear filters—one for each prediction horizon—is constructed. All filters share the same input vector \( \mathbf{x}_k \) defined in~\eqref{eq:input_vector}. The \( \tau \)-step SINR prediction is expressed as a linear combination of past samples:
\begin{equation}
	\label{eq:WienerOutput}
	\hat{\gamma}_{\mathrm{e}}(T_\mathrm{CSI} + \tau) = \mathbf{a}_{\tau}^\top \mathbf{x}_{k},
\end{equation}
where \( \mathbf{a}_{\tau} \in \mathbb{R}^{P \times 1} \) are the filter coefficients for horizon \( \tau \) and \((\cdot)^\top\) denotes the transpose. The optimal coefficients are found by solving the Wiener–Hopf equations:
\begin{equation}
	\mathbf{a}_{\tau} = \mathbf{R}^{-1}\mathbf{r}_{\tau},
\end{equation}
where \( \mathbf{R}=\mathbb{E}\left[\mathbf{x}_k \cdot \mathbf{x}_k^\top \right] \in \mathbb{R}^{P \times P} \) is the autocorrelation matrix of the input vector, whose elements can be expressed in terms of the autocorrelation function of \(\gamma_{\mathrm{e}}(n)\), \(R_{\gamma_{\mathrm{e}}}(m)\), as
\begin{equation} \label{eq:autocorrelationExp}
	[\mathbf{R}]_{i,j} = \mathbb{E} \!\left[ \gamma_{\mathrm{e}}(T_\mathrm{CSI} \cdot i) \, \gamma_{\mathrm{e}}(T_\mathrm{CSI} \cdot j) \right] = R_{\gamma_{\mathrm{e}}}(T_{\mathrm{CSI}} \cdot (i-j)),
\end{equation}
with $i,j \in [0, P-1]$. The vector \( \mathbf{r}_{\tau} \in \mathbb{R}^{P \times 1} \) is the cross-correlation between the input vector and the desired future value:
\begin{equation}
    \mathbf{r}_{\tau} = \mathbb{E}\!\left[ \mathbf{x}_k \gamma_{\mathrm{e}}(T_{\mathrm{CSI}}k + \tau) \right],
\end{equation}
whose elements can be also expressed in terms of the autocorrelation function of \(\gamma_{\mathrm{e}}(n)\) as
\begin{equation}
	[\mathbf{r}_{\tau}]_i = \mathbb{E} \!\left[ \gamma_{\mathrm{e}}(T_\mathrm{CSI} (k-i)) \gamma_{\mathrm{e}}(T_\mathrm{CSI} k + \tau) \right] = R_{\gamma_{\mathrm{e}}}(-T_{\mathrm{CSI}}i+\tau).
\end{equation}

The prediction error for horizon \( \tau \) is given by
\begin{equation}
	e(T_\mathrm{CSI} k + \tau) = \gamma_{\mathrm{e}}(T_\mathrm{CSI} k + \tau) - \hat{\gamma}_{\mathrm{e}}(T_\mathrm{CSI} k + \tau),
\end{equation}
and its mean-square value (the minimum MSE) is
\begin{equation}
    \label{eq:errorNoPrediction}
	\mathbb{E}[|e(T_\mathrm{CSI} k + \tau)|^2] = \sigma_{\gamma_{\mathrm{e}}}^2
	- \mathbf{r}_{\tau}^\top\mathbf{R}^{-1} \,\mathbf{r}_{\tau} ,
\end{equation}
where \(\sigma_{\gamma_{\mathrm{e}}}^2\) denotes the variance of \(\gamma_{\mathrm{e}}(n)\). 

It is well known that as the prediction horizon increases ($\tau \to \infty$), the mean squared error of the Wiener predictor approaches the signal variance, $\sigma_{\gamma_{\mathrm{e}}}^2\)~\cite[Sec.~5.7]{Brockwell1987}. Therefore, even for long horizons, the Wiener filter outperforms a $\tau$-sample zero-order hold (ZOH) predictor, whose error power converges to $2\sigma_{\gamma_{\mathrm{e}}}^2$.

Since the second-order statistic of the process, \(R_{\gamma_{\mathrm{e}}}(m)\), is unknown, it is estimated empirically from training data under the assumption of ergodicity, so that time averages approximate ensemble expectations.

The computational complexity of the Wiener filter's inference phase is defined by $P$, the filter order. Once the $P$ coefficients of a filter $\mathbf{a}_{\tau}$ are known, the prediction $\hat{\gamma}_{\mathrm{e}}(T_\mathrm{CSI} + \tau)$ is generated by (6). This equation represents a single dot product. This operation for a \textit{single filter} requires $P$ multiplications and $P-1$ additions (totaling $2P-1$ FLOPs). Since a bank of $(T_{\mathrm{CSI}}-1)$ filters is used (one for each prediction step $\tau$), the total number of \ac{FLOPs} for the complete inference phase is 
\begin{equation}
	\mathrm{FLOPs}_{\mathrm{Wiener}_{\mathrm{bestCQI}}} = (T_{\mathrm{CSI}}-1) \cdot (2P-1),
\end{equation}
for the \textbf{by-CQI} case, where a prediction is made for each of the $N_{\mathrm{CQI}}$ levels, the total complexity is $N_{\mathrm{CQI}}$ times greater:
\begin{equation}
	\mathrm{FLOPs}_{\mathrm{Wiener}_{\mathrm{byCQI}}} = N_{\mathrm{CQI}}\cdot \mathrm{FLOPs}_{\mathrm{Wiener}_{\mathrm{bestCQI}}}.
\end{equation}

\subsection{AI-Based Prediction Networks}
\begin{figure}[htbp]
    \centering
    \includegraphics[width=0.6\linewidth]{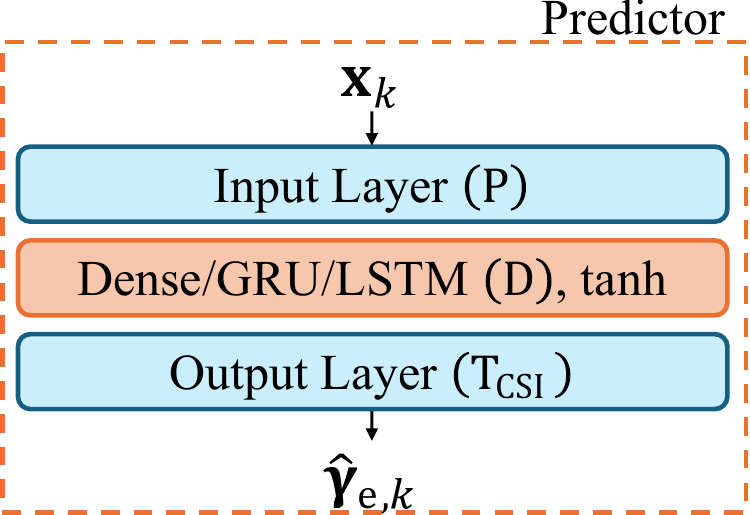}
    \caption{Unified architecture of the proposed AI-based prediction networks.}
    \label{fig:PredictorIA}
\end{figure}

The proposed AI-based \ac{CSI} prediction framework adopts the unified structure illustrated in Fig.~\ref{fig:PredictorIA}, enabling a fair comparison among different learning architectures in terms of accuracy and computational complexity. It comprises three main components: 1) an \textbf{input layer} that receives a sequence of $P$ past effective SINR samples; 2) a \textbf{hidden layer} that captures temporal or nonlinear dependencies using Dense, \ac{GRU}, or \ac{LSTM} units with a \textit{tanh} activation function; and 3) an \textbf{output layer} that predicts the required effective \ac{SINR} values.

\subsubsection{DNN-Based Prediction Network}
As a baseline, we implement a fully connected \ac{DNN} with a single hidden layer that learns a nonlinear mapping between past and future SINR values. At each reporting instant \( k \), the input is the vector \( \mathbf{x}_k \) from~\eqref{eq:input_vector}. The hidden layer computes its output \( \mathbf{h} \) as
\begin{equation}
    \mathbf{h}_k = \sigma \left( \mathbf{W}^{(1)} \mathbf{x}_k + \mathbf{b}^{(1)} \right),
\end{equation}
where \( \sigma(\cdot) \) denotes a nonlinear activation function (e.g., ReLU), and \( \mathbf{W}^{(1)}, \mathbf{b}^{(1)} \) are the weights and biases. Since the input layer uses past samples of the input sequence $\gamma_\mathrm{e}(n)$ we named this network delayed \ac{DNN}. The output layer then produces the predicted SINR vector
\begin{equation}
    \hat{\boldsymbol{\gamma}}_{\mathrm{e},k} = 
    \mathbf{W}^{(2)} \mathbf{h}_k + \mathbf{b}^{(2)}.
\end{equation}
The model is trained by minimizing the MSE loss function
\begin{equation}
    \mathcal{L} = \frac{1}{T_{\mathrm{CSI}}-1} \sum_{\tau=1}^{T_{\mathrm{CSI}}-1} 
    \left( \gamma_{\mathrm{e}}(T_{\mathrm{CSI}} k+\tau) - \hat{\gamma}_{\mathrm{e}}(T_{\mathrm{CSI}} k+\tau)\right)^2.
\end{equation}


The computational complexity of the inference phase for the \ac{DNN} is determined by the operations in its hidden and output layers. Let $P$ be the input sequence length and $D$ be the number of units in the hidden layer. The hidden layer requiring $2PD$ FLOPs. The output layer then takes the hidden activation and projects it to the output size of $T_{\mathrm{CSI}}-1$, which requires $2D(T_{CSI}-1)$ FLOPs.

The total complexity for the \textbf{best-CQI} case is the sum of both:
\begin{equation}
    \mathrm{FLOPs}_{\mathrm{DNN}_{\mathrm{best-CQI}}} =  2D(P + T_{\mathrm{CSI}} - 1), 
\end{equation}
for the \textbf{by-CQI} case, where a prediction is made for each of the $N_{\mathrm{CQI}}$ levels, the total complexity is $N_{\mathrm{CQI}}$ times greater:
\begin{equation}
    \mathrm{FLOPs}_{\mathrm{DNN}_{\mathrm{by-CQI}}} = N_{\mathrm{CQI}} \cdot \mathrm{FLOPs}_{\mathrm{DNN}_{\mathrm{best-CQI}}}.
\end{equation}

\subsubsection{LSTM-Based Prediction Network}
We also develop a \ac{RNN} using \ac{LSTM} units to capture long-term temporal dependencies in the channel. 


\begin{figure}[htbp]
    \centering
    \includegraphics[width=0.9\linewidth]{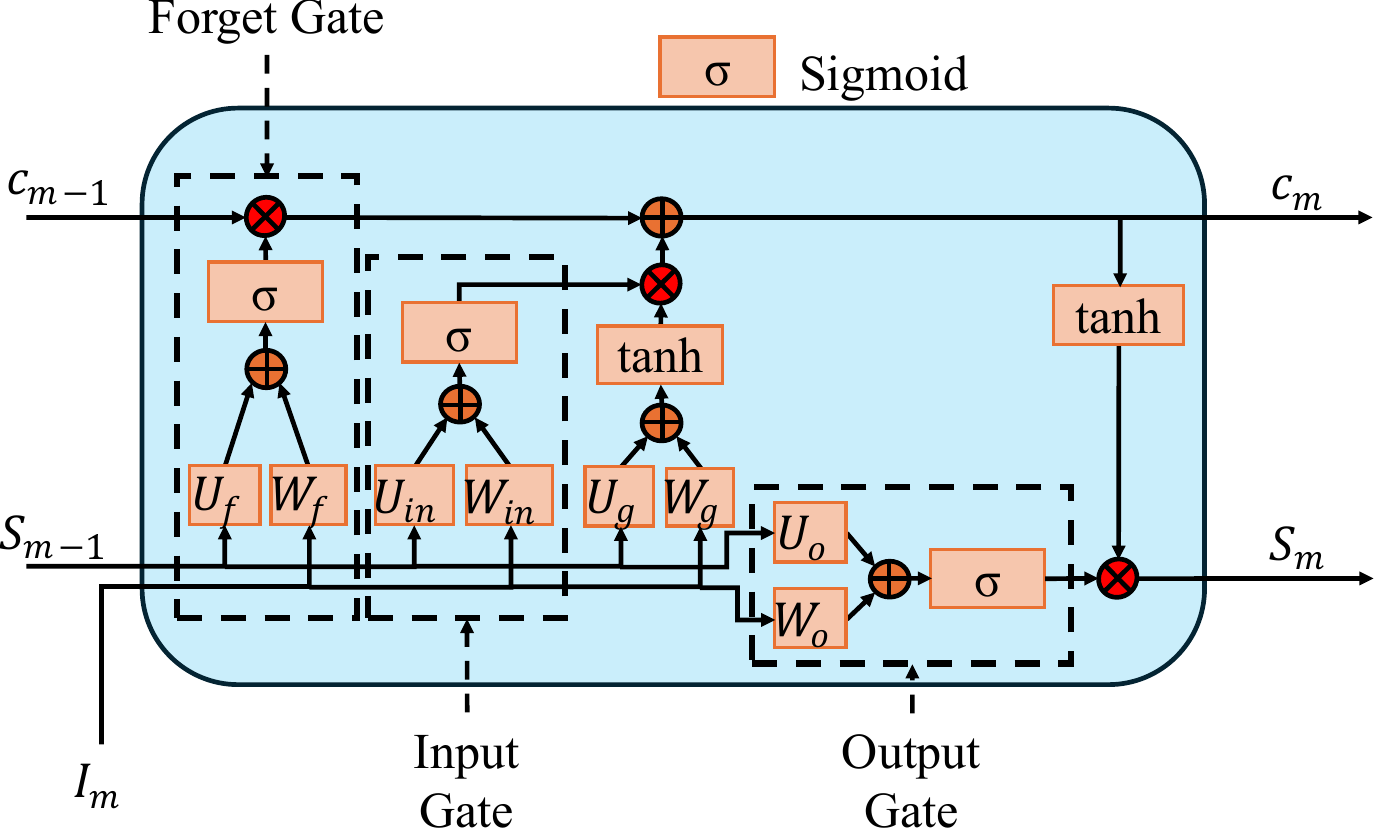}
    \caption{Internal structure of an \ac{LSTM} unit.}
    \label{fig:LSTMCell}
\end{figure}

As shown in Fig.~\ref{fig:LSTMCell}, each \ac{LSTM} cell comprises three gating mechanisms—the input, forget, and output gates—each governed by a sigmoid activation that outputs values in the range \([0,1]\). These gates regulate the flow of information by controlling how much of the input is retained, updated, or discarded at each time step. During training, the network adjusts the associated weight matrices (\(\mathbf{U}_x\) and \(\mathbf{W}_x\)) to emphasize relevant temporal patterns, producing gate activations closer to one for informative features~\cite{Mattu2022}.

At each time step \( m \), the \ac{LSTM} cell processes an input sample \( I_m \) and maintains two internal state vectors: the hidden state vector \( \mathbf{S}_m \in \mathbb{R}^D \), representing the cell's output at that step, and the cell state vector \( \mathbf{c}_m \in \mathbb{R}^D \), which acts as a memory channel to preserve long-term dependencies. These states are dynamically updated through the interaction of the gating mechanisms, allowing the \ac{LSTM} to retain or discard information based on its learned temporal relevance.

A key advantage of \ac{LSTM} over standard \ac{RNN} architectures is their ability to learn long-term correlations while mitigating the vanishing gradient problem~\cite{Hochreiter1997LSTM}, making them highly suitable for capturing the complex dynamics of wireless channels in \ac{CSI} prediction tasks.

Each LSTM cell (Fig.~\ref{fig:LSTMCell}) contains four gate structures (forget, input, output, and candidate). We calculate the FLOPs per step by adding the cost of these gates and the subsequent elemental operations.

The four gates contribute a total of $8D^2 + 8D$ FLOPs. The elemental operations to update the cell state and hidden state add additional $4D$ FLOPs. The total FLOPs per time step is $8D^2 + 12D$.

The total complexity for the \textbf{best-CQI} case is the sum of $P$ recurrent steps plus the linear output layer (which costs $2D(T_{\mathrm{CSI}}-1)$ FLOPs):
\begin{equation}
    \mathrm{FLOPs}_{\mathrm{LSTM}_{\mathrm{best-CQI}}} = P(8D^2 + 12D) + 2D(T_{\mathrm{CSI}} - 1),
\end{equation}
for the \textbf{by-CQI} scenario, this entire complexity is scaled by the $N_{\mathrm{CQI}}$ levels:
\begin{equation}
    \mathrm{FLOPs}_{\mathrm{LSTM}_{\mathrm{by-CQI}}} = N_{\mathrm{CQI}} \cdot \mathrm{FLOPs}_{\mathrm{LSTM}_{\mathrm{best-CQI}}}.
\end{equation}
 
\subsubsection{GRU-Based Prediction Network}
As an alternative to the \ac{LSTM}, we implement a prediction network based on \ac{GRU}. The \ac{GRU} architecture offers a simplified yet powerful recurrent structure by merging the input and forget gates into a single ``update gate'' and combining the cell state and hidden state. This streamlined design reduces parameters and computations, maintaining comparable temporal modeling capabilities~\cite{chung2014empiricalevaluationgatedrecurrent}.

Similar to the \ac{LSTM}, the \ac{GRU}-based predictor consists of an input layer, a hidden layer with \ac{GRU} units, and an output layer that generates the required effective \ac{SINR} values. Due to its reduced parameter count, the \ac{GRU} typically exhibits faster training times and lower inference latency compared to an \ac{LSTM} of similar capacity.

The \ac{GRU} simplifies the \ac{LSTM} architecture, reducing computational complexity. It consists of two main gates and a candidate state calculation, totaling three structures that perform matrix operations.

The three gate-like structures contribute a total of $6D^2 + 7D$ FLOPs, and the final elemental state update adds $4D$ FLOPs. The total FLOPs per time step sums to $6D^2 + 11D$.

The total complexity for inference in the \textbf{best-CQI} case is the sum of $P$ recurrent steps plus the final linear layer:
\begin{equation}
    \mathrm{FLOPs}_{\mathrm{GRU}_{\mathrm{best-CQI}}} = P \cdot (6D^2 + 11D) + 2D(T_{\mathrm{CSI}} - 1),
\end{equation}
for the \textbf{by-CQI} case, the complexity scales linearly with the number of CQI levels:
\begin{equation}
    \mathrm{FLOPs}_{\mathrm{GRU}_{\mathrm{by-CQI}}} = N_{\mathrm{CQI}} \cdot \mathrm{FLOPs}_{\mathrm{GRU}_{\mathrm{best-CQI}}}.
\end{equation}

\section{Numerical Results}

This section investigates the performance of the proposed \ac{CSI} prediction frameworks under different propagation conditions and evaluating a diverse set of key performance indicators. The first subsection focuses on model design, analyzing how the model complexity, Doppler shift, and the \ac{CSI} reporting period influence prediction accuracy for the Wiener predictor and the presented AI-based models, including \ac{LSTM}, \ac{GRU} and delayed DNN. The second subsection presents end-to-end system results, comparing the performance of the \ac{GRU}-based predictor with that of the classical Wiener filter. Additionally, we examine the generalization capabilities of both approaches under diverse channel conditions, including variations in delay profiles and Doppler frequencies. The simulation parameters used throughout this study are summarized in Table~\ref{tab:SimulationParameters}.

\begin{table}[htbp]
    \centering
    \caption{Simulation Parameters}
    \label{tab:SimulationParameters}
    \begin{tabular}{lc}
        \toprule
        \textbf{Parameter} & \textbf{Value} \\
        \midrule
        Average SNR & 12.5 dB \\
        Subcarrier Spacing (SCS) & 15 kHz \\
        Number of RBs ($N_{\text{RB}}$) & 52 \\
        Bandwidth & 10 MHz \\
        Doppler Shift ($f_{D}$) & 1--50 Hz \\
        Delay Spread & 300 ns \\
        MIMO Configuration & $4 \times 4$ \\
        Transmission Layers & 4 \\
        Channel Models & TDL-A, B, C, D, E \\
        CSI Reporting Period ($T_{\text{CSI}}$) & \{4, 32, 40\} slots \\
        Transmission Interval & 1 ms \\
        \bottomrule
    \end{tabular}
\end{table}

The datasets used for training and evaluation were generated using a custom link-level simulator developed in MATLAB. This simulator emulates a \ac{MIMO}-\ac{OFDM} wireless system operating under standardized 3GPP conditions and supports multiple \ac{TDL} channel profiles, including TDL-A, TDL-B, TDL-C, TDL-D, and TDL-E~\cite{38.901}. These profiles characterize a broad range of propagation environments, with TDL-A, TDL-B, and TDL-C representing varying degrees of \ac{NLOS} conditions, and TDL-D and TDL-E corresponding to \ac{LOS} scenarios. 

Each dataset was generated for specific simulation parameters, including Doppler frequencies ranging from 1~Hz to 50~Hz and a delay spread of 300~ns. The simulator computes the time-varying channel response over multiple slots and calculates the corresponding effective SINR values using the EESM method. The resulting SINR sequences are then used to construct input-output pairs for training and testing the prediction models. A total of 510,000 samples were generated, divided into 78.4\% for training, 19.6\% for validation, and 2\% for testing. Both models were trained for 200 epochs with a batch size of 2048 to ensure convergence across diverse channel conditions.


\subsection{Model design}
We present a series of empirical tests performed to identify the optimal configuration of the proposed prediction models. Here, we describe the experimental design, such as dependence with Doppler and CSI reporting period, prediction based on real data, and study of the optimal model complexity. 

\subsubsection{Trade-off Between Performance and Complexity} This subsection analyzes the trade-off between prediction accuracy and computational complexity across different model architectures and identifies the optimal input sequence length for the proposed predictors. The goal is to determine a configuration that achieves high accuracy while maintaining low inference cost.

\begin{figure}[htbp]
    \centering
    \includegraphics[width=0.9\linewidth]{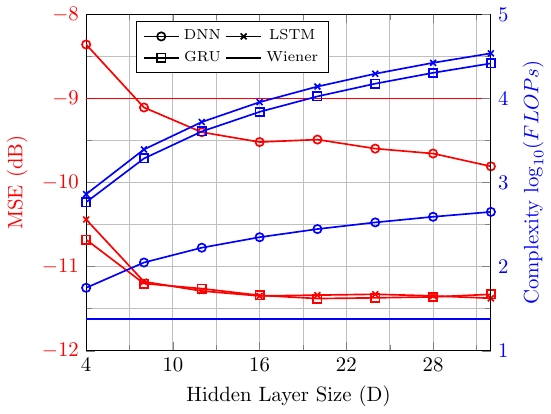}
    \caption{Trade-off between MSE (dB) and computational complexity (FLOPs) for different prediction frameworks as a function of hidden-layer size, for a Doppler frequency of 10~Hz and \( T_{\mathrm{CSI}}=4 \) slots in the TDL-A channel.}
    \label{fig:ComplexityComparation}
\end{figure}

Fig.~\ref{fig:ComplexityComparation} illustrates the relationship between model complexity and prediction accuracy (measured in MSE) for all evaluated frameworks. As the hidden-layer size increases, the MSE decreases, but the gain saturates beyond moderate model capacities. Among the learning-based approaches, both the \ac{LSTM} and \ac{GRU} achieve the lowest MSE values, outperforming the DNN and Wiener filter. However, the \ac{GRU} offers a lower computational cost that of the \ac{LSTM} while maintaining nearly identical accuracy. The DNN shows limited temporal modeling capability, and the Wiener filter provides the simplest implementation, requiring only a few FLOPs per inference at the expense of slightly higher MSE. 

Based on these results, a \ac{GRU} with \(D=16\) hidden units is selected as the representative learning-based predictor for subsequent analyses, offering prediction accuracy comparable to the \ac{LSTM} but with a lower computational cost. This configuration entails a complexity of approximately 6944 FLOPs. The Wiener filter is retained as a low-complexity benchmark, providing a fair basis for comparing classical and AI-based prediction strategies. We investigate the impact of model depth and input sequence length on prediction performance, summarizing the key observations in the following remark.

\begin{remark}[Optimal Model Complexity]
At higher Doppler frequencies, increasing model depth or input sequence length yields marginal or no improvement in prediction accuracy. Therefore, compact models achieve comparable performance with substantially lower computational requirements.
\end{remark} 

To further investigate the influence of input length on prediction performance, we evaluated the MSE obtained by \ac{GRU}- and Wiener-based predictors for different sequence sizes and Doppler frequencies. For the \ac{GRU}, the input length corresponds to the number of past effective SINR samples provided to the network, whereas in the Wiener filter it represents the filter order. Table~\ref{tab:NMSEComparisonJoint} summarizes the results. Accuracy improves initially as the input window expands but saturates beyond four samples for both models. At higher Doppler frequencies, longer input sequences can even degrade performance due to redundancy or overfitting.

\begin{table}[h]
    \caption{MSE (dB) versus input sequence length (for \ac{GRU}) and filter order (for Wiener) under different Doppler frequencies, with \( T_{\mathrm{CSI}} = 4 \) slots.}
    \centering
    \renewcommand{\arraystretch}{1.2}
    \setlength{\tabcolsep}{3pt}
    \begin{tabular}{c|c|c|c|c|c|c|c|c}
        \hline
        & \multicolumn{8}{c}{\textbf{Input Size / Filter Order}} \\
        \hline
        \textbf{Model} & \textbf{$f_{D}$} & 1 & 2 & 3 & 4 & 5 & 6 & 7 \\
        \hline
        \multirow{3}{*}{\textbf{GRU}} 
        & 10 Hz & -7.8 & -9.7 & -10.5 & \textbf{-11.2} & -11.5 & -11.4 & -11.4 \\
        & 20 Hz & -3.4 & -4.7 & \textbf{-5.0} & -5.0 & -5.0 & -5.0 & -5.0 \\
        & 40 Hz & -0.5 & \textbf{-0.7} & -0.6 & -0.6 & -0.7 & -0.6 & -0.6 \\
        \hline
        \multirow{3}{*}{\textbf{Wiener}} 
        & 10 Hz & -7.7 & -8.9 & -8.9 & \textbf{-9.0} & -9.0 & -9.0 & -9.0 \\
        & 20 Hz & -3.3 & -4.1 & \textbf{-4.2} & -4.2 & -4.2 & -4.2 & -4.2 \\
        & 40 Hz & -0.53 & \textbf{-0.6} & -0.6 & -0.6 & -0.6 & -0.6 & -0.6 \\
        \hline
    \end{tabular}
    \label{tab:NMSEComparisonJoint}
\end{table}
Based on these observations, an input length (or filter order) of four provides an effective trade-off between accuracy and complexity for both predictors. This configuration is adopted in the subsequent performance evaluations, ensuring robust and efficient operation across diverse channel conditions.

\subsubsection{Dependence on Doppler and \ac{CSI} reporting period}

The prediction accuracy is significantly influenced by the channel's temporal dynamics (Doppler frequency) and the periodicity of \ac{CSI} acquisition. In prediction tasks, these two factors jointly affect the difficulty of accurately forecasting future channel states. This subsection investigates the impact of the $f_D \times T_{\mathrm{CSI}}$ product on prediction performance.

\begin{remark}[Dependence on Doppler and CSI reporting period]
Simulation results reveal that the performance of CSI prediction, as measured by MSE, is primarily determined by the product \( f_D \times T_{\mathrm{CSI}} \). That is, different combinations of Doppler frequency \( f_D \) and CSI reporting interval \( T_{\mathrm{CSI}} \) that yield the same product result in similar prediction accuracy.
\end{remark}

\begin{figure}[htbp]
    \centering
    \includegraphics[width=0.9\linewidth]{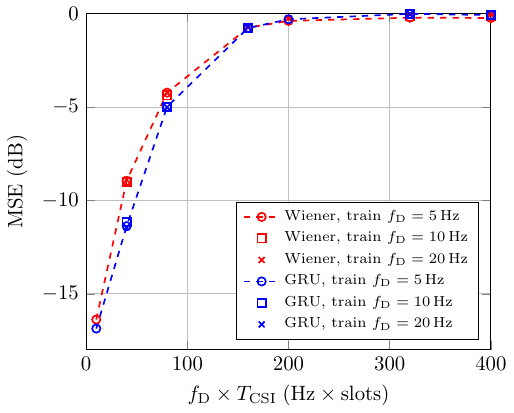}
    \caption{MSE (dB) performance of GRU and Wiener-based CSI predictors for different values of the product \( f_{\mathrm{D}} \times T_{\mathrm{CSI}} \) in the TDL-A channel. Each curve corresponds to a model trained at a specific Doppler frequency (\( f_{\mathrm{D}} = 5 \), 10, and 20 Hz) and evaluated at prediction horizons \( \tau \in \{1, 4, 8, 16, 32, 40, 64, 80\} \) slots.}
    \label{grap:FDTcsiDependance}
\end{figure}

Fig.~\ref{grap:FDTcsiDependance} illustrates the prediction error in terms of MSE for both \ac{GRU} and Wiener-based predictors, as a function of the product \( f_D \times T_{\mathrm{CSI}} \). Each curve corresponds to a model trained at a specific Doppler frequency (\( f_D = 5 \), 10, and 20 Hz), and evaluated across multiple values of \( T_{\mathrm{CSI}} \) to maintain different values of the product.

The results exhibit a clear dependency of prediction performance on the value of \( f_D \times T_{\mathrm{CSI}} \), supporting the hypothesis that this product effectively captures the temporal variability of the channel. Notably, for small values of \( f_D \times T_{\mathrm{CSI}} \) (i.e., low mobility and/or high CSI reporting rate), both predictors achieve very low MSE values, with the \ac{GRU} predictor reaching performance levels close to \( -17 \,\mathrm{dB} \), indicating highly accurate predictions.

However, as the product increases, the prediction error also increases. 
This degradation follows a distinctive ``waterfall'' behavior: a relatively sharp transition occurs around \( f_D \times T_{\mathrm{CSI}} \approx 150\text{--}200 \), beyond which the MSE rapidly deteriorates from highly accurate to values close to 0 dB. 

Importantly, the region where the MSE remains below \( -10 \,\mathrm{dB} \) can be regarded as the operational range where the predicted CSI is sufficiently accurate for practical use in link adaptation. For \( f_D \times T_{\mathrm{CSI}} > 200 \), the MSE approaches the variance of the signal itself, thus leading to MSE $\simeq 0$ dB. 



\subsubsection{Processing of Past Samples}
An important design decision in \ac{CSI} prediction frameworks concerns how historical observations are processed at the model input. Reference signals are transmitted periodically, so the network input is available only at the slots where these signals occur. To address this, we consider two distinct approaches: \textbf{interpolated inputs}, where estimation techniques are applied to fill the missing slots and densify the sequence, and \textbf{non-interpolated inputs}, which utilize the original, sparsely sampled data directly. Although interpolation intuitively suggests a richer input, it is unclear whether this improves prediction accuracy compared to using the raw sparse data. Therefore, the impact of such preprocessing is evaluated empirically.

\begin{remark}[Input sampling] Simulation results indicate that using actual, non-interpolated input samples yields better prediction performance than using interpolated inputs, particularly for higher values of \( f_D \times T_{\mathrm{CSI}} \).
\end{remark}

\begin{figure}[htbp]
    \centering
    \includegraphics[width=0.9\linewidth]{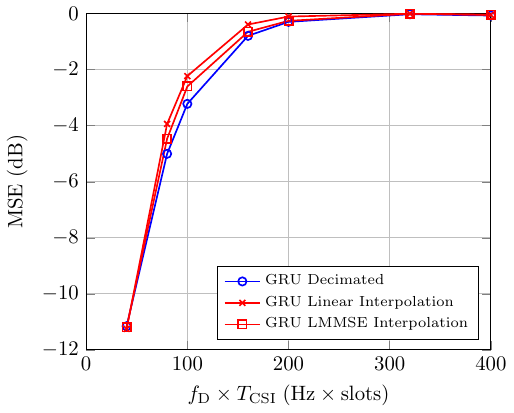}
    \caption{Comparison of prediction error in terms of MSE (dB) between a GRU model using interpolated inputs (linear and LMMSE) and one using actual reference measurements only, evaluated in the TDL-A channel with \( f_{\mathrm{D}} = 10~\mathrm{Hz} \) at prediction horizons \( \tau \in \{4, 8, 10, 16, 20, 32, 40\} \) slots.}
    \label{grap:InterpolationVSDecimated}
\end{figure}

Fig.~\ref{grap:InterpolationVSDecimated} compares the prediction error, expressed in MSE, for two \ac{GRU}-based models: one using decimated input sequences composed only of actual pilot-based measurements spaced by \( T_{\mathrm{CSI}} \), and another using interpolated inputs obtained via piece-wise linear interpolation and linear MMSE (LMMSE) methods \cite{kay1993fundamentals}.

At low values of \( f_D \times T_{\mathrm{CSI}} \) (i.e., under slow channel variations or when predicting a short horizon into the future) the performance of all methods is comparable. This is expected as the channel exhibits limited variation between reference points, and simple interpolation suffices. However, as \( f_D \times T_{\mathrm{CSI}} \) increases, the advantage of the decimated model becomes more evident. In these cases, the \ac{GRU} network is able to perform a form of non-linear, interpolation that better captures the complex dynamics of the channel, outperforming classical linear methods.

This behavior highlights a key distinction: when interpolation is performed explicitly as a pre-processing step, it may introduce smoothing or approximation artifacts that degrade prediction quality. In contrast, allowing the \ac{GRU} to operate directly on sparse, real observations enables the model to learn and apply its own implicit interpolation strategy, which is more robust under high channel variability. Additionally, the decimated approach reduces computational overhead by avoiding interpolation altogether.

\subsubsection{Prediction Target Strategy: Best-CQI vs By-CQI}

An important design consideration when training CSI prediction models is the choice of the prediction target. Two main strategies are compared: (i) \textbf{By-CQI}, which predicts the effective SINR values for all possible CQI levels (i.e., a full CQI vector), and (ii) \textbf{Best-CQI}, which predicts only the effective SINR corresponding to the best CQI selected during link adaptation.

\begin{remark}
The simulation results show that the Best-CQI technique exhibits a higher prediction error compared to the By-CQI technique. However, this improvement in accuracy with By-CQI entails a computational complexity increased by a factor of \( N_{\mathrm{CQI}} = 15 \).
\end{remark}

\begin{figure}[htbp]
    \centering
    \includegraphics[width=0.9\linewidth]{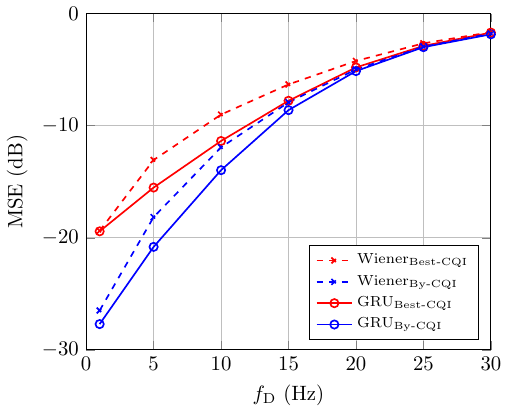}
    \caption{MSE (dB) comparison between two prediction target strategies Best-CQI vs. by-CQI for both Wiener and GRU predictors in the TDL-A channel, with $T_{\mathrm{CSI}}=4$ slots and $f_{\mathrm{D}} = 10~\mathrm{Hz}$.}
    \label{grap:BestCQIvsbyCQI}
\end{figure}

Fig.~\ref{grap:BestCQIvsbyCQI} illustrates the trade-off introduced by the Best-CQI strategy. While this approach reduces model complexity $N_\mathrm{CQI}=15$ times by focusing on the effective \ac{SINR} corresponding to the selected CQI level, it results in lower prediction accuracy compared to the By-CQI strategy. In high-mobility scenarios, where the channel varies rapidly, prediction performance degrades for both techniques. In contrast, in low-mobility scenarios, where the channel evolves more slowly, there is a greater opportunity for accurate prediction, and the advantages of the By-CQI strategy become more evident. This highlights the trade-off between computational efficiency and prediction accuracy, which must be carefully considered depending on the mobility characteristics of the target environment.

\subsection{Model evaluation}
In this section, we evaluate the performance of the proposed CSI prediction frameworks in practical link adaptation scenarios. 
The evaluation addresses \ac{TDD} and \ac{FDD} systems separately due to their distinct \ac{CSI} requirements.

\subsubsection{CSI acquisition in TDD}

This subsection demonstrates how the use of a prediction algorithm for CQI can enhance end-to-end performance in terms of throughput in TDD systems. It also evaluates the proposed CSI prediction models considering their generalizability under various scenarios. 


\textbf{End-to-end results:} 
%
%
We compare below the performance of both the \ac{GRU} and Wiener-based predictors in terms of downlink throughput. 

\begin{figure}[htbp]
    \centering
    \includegraphics[width=0.9\linewidth]{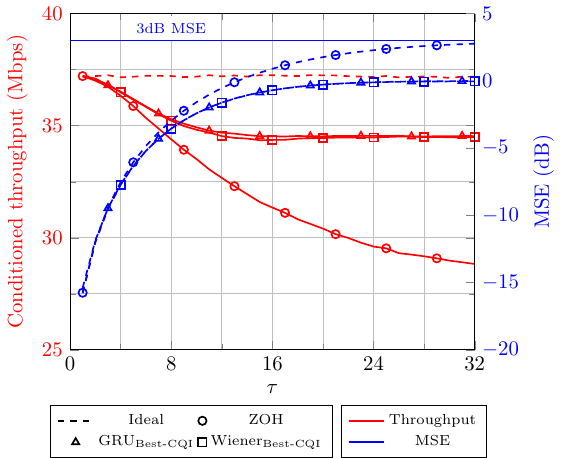}
    \caption{Comparison of conditioned throughput in Mbps (left y-axis) and MSE in dB (right y-axis) for different CSI acquisition strategies. Results are averaged over multiple channel realizations with \( T_{\mathrm{CSI}} = 32 \) slots and \( f_D = 10 \) Hz.}
    \label{fig:ThroughputAndNMSE}
\end{figure}

Fig.~\ref{fig:ThroughputAndNMSE} illustrates the trade-off between throughput and prediction accuracy for various CSI acquisition strategies. The x-axis (\( \tau \in [1, T_\mathrm{CSI}-1] \)) represents the slot index within the CSI reporting interval, indicating the delay lag of each throughput or prediction error measurement relative to the last CSI information that was acquired (see Fig. \ref{fig:transmissionInterval}). The left y-axis shows the average downlink throughput, while the right y-axis reports the MSE  associated with each prediction method. Four configurations are evaluated in terms of throughput: ideal CSI, ZOH, \ac{GRU}-based prediction, and Wiener-based prediction. The MSE curves are only shown for the non-ideal schemes, since the ideal case assumes perfect channel knowledge and therefore has zero prediction error. The ideal CSI scenario achieves the highest throughput, serving as an upper bound.

The ZOH scheme shows notable degradation due to the outdated CSI, leading to mismatches in the selection of MCS. Both prediction-based schemes improve throughput over this baseline, confirming the value of anticipating channel variations. For small slot indices (e.g. \( \tau < 6 \)), the ZOH strategy performs comparably well, making the prediction unnecessary. However, for larger \( \tau \), both prediction methods significantly outperform the baseline ZOH. In this regime, the Wiener and \ac{GRU} predictors achieve nearly identical throughput, with only a marginal difference.

In addition, Fig.~\ref{fig:ThroughputAndNMSE} highlights that \textit{applying the prediction, even with a MSE around 0~dB, leads to a higher throughput (and smaller MSE) than not predicting at all (i.e., ZOH)}. As discussed in section \ref{subsec:WienerFilter} and \cite[Sec.~5.7]{Brockwell1987}, this is due to the accumulation of prediction error in the \textit{ZOH} case as the horizon increases. As indicated below \ref{eq:errorNoPrediction}, when predicting \( \tau \to \infty \) slots ahead without updating the CSI, the normalized error power converges to approximately 3~dB.

\begin{figure}[htbp]
    \centering
    \subfloat[]{
        \includegraphics[width=0.9\linewidth]{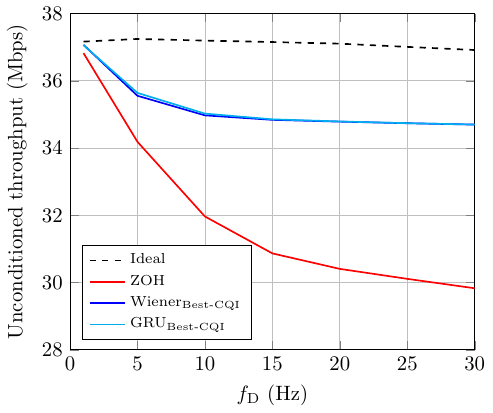}
        \label{fig:throughputTcsi32}
    }
    \\
    \subfloat[]{
        \includegraphics[width=0.9\linewidth]{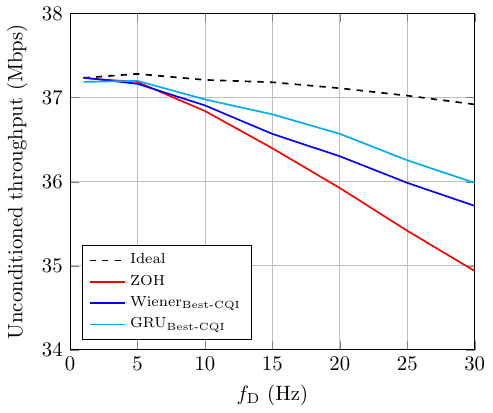}
        \label{fig:throughputTcsi4}
    }
    \caption{Unconditioned throughput as a function of Doppler frequency for different CSI prediction strategies in the TDL-A channel, considering reporting intervals of (a) \( T_{\mathrm{CSI}} = 32 \) slots and (b) \( T_{\mathrm{CSI}} = 4 \) slots.}
    \label{fig:throughput_comparison}
\end{figure}

Fig.~\ref{fig:throughput_comparison} analyzes the unconditioned throughput as a function of Doppler frequency \( f_D \) for different reporting intervals. As shown in Fig.~\ref{fig:throughput_comparison}\subref{fig:throughputTcsi32}, using a \ac{CSI} reporting interval of \( T_{\mathrm{CSI}} = 32 \) slots amplifies the performance gap between prediction-based schemes and the ZOH baseline. While the throughput gain from prediction is negligible at low Doppler frequencies (e.g., \( f_D = 1 \)~Hz), the benefit becomes substantial as \( f_D \) increases and the channel varies more rapidly. In this setting, both prediction algorithms significantly improve unconditioned throughput compared to the non-predictive approach. On average, the \ac{GRU}-based predictor achieves a throughput gain of 10.5\%, while the Wiener-based predictor provides a comparable gain of 10.43\%.

Conversely, Fig.~\ref{fig:throughput_comparison}\subref{fig:throughputTcsi4} examines the scenario with a frequent reporting interval of \( T_{\mathrm{CSI}} = 4 \) slots. When Doppler is low, the channel varies slowly between reports, resulting in similar performance for ZOH, Wiener, and \ac{GRU}. However, as \( f_D \) increases, prediction-based schemes progressively outperform the ZOH baseline, with the \ac{GRU} model consistently achieving slightly higher throughput than the Wiener predictor. Quantitatively, the gains are more modest due to the quasi-static nature of the channel between frequent updates: at \( f_D = 30 \)~Hz, the \ac{GRU}-based predictor achieves a throughput gain of 2.79\%, while the Wiener-based predictor yields a gain of 2.21\%, relative to the ZOH case.

Crucially, this frequent reporting regime (\( T_{\mathrm{CSI}} = 4 \)) incurs high signaling overhead and is precisely what efficient system design aims to avoid. The core value of \ac{CSI} prediction lies in enabling the use of longer reporting intervals (such as \( T_{\mathrm{CSI}} = 32 \)) to drastically reduce overhead while maintaining high throughput, a balance that is unachievable with standard ZOH techniques.

\textbf{Generalization capabilities:} An important requirement for practical CSI prediction frameworks is the ability to generalize across different channel conditions without retraining. In this subsection, we analyze and compare the generalization capabilities of the \ac{GRU} and Wiener-based predictors.

First, we assess how well the models designed/trained for a specific Doppler frequency \( f_D \) perform when tested under the same Doppler condition (model switch case) versus models trained using a mixture of Doppler values (generalized case). This allows us to evaluate whether a single generalized model can replace multiple Doppler-specific models without compromising performance. This latter approach is known as \textbf{model life-cycle management} and it considers that the network has different models that have been trained for different conditions \cite{3gpp_ran112}. Then, the network monitors the performance of the currently deployed model, and it might require a model switch to change the model that is being deployed for inference. Therefore, the best model can be used for the current channel conditions. 

Second, we extend the evaluation to different channel models by testing both the specific and generalized models over various TDL profiles, including TDL-A through TDL-E. This step quantifies the robustness of each predictor to variations in delay spread and multipath richness, which are critical factors in real-world wireless environments.

\begin{figure}[htbp]
    \centering
    \includegraphics[width=0.9\linewidth]{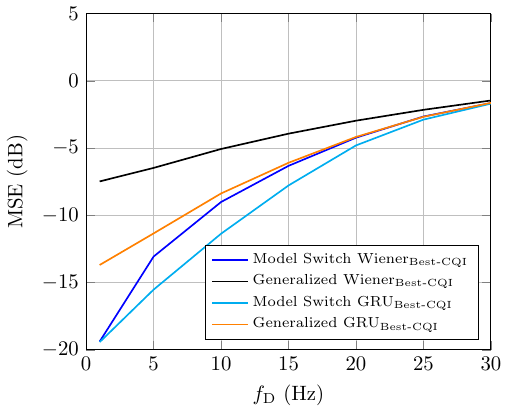}
    \caption{MSE performance of Wiener and GRU models under different Doppler shifts, comparing model-specific and generalized training strategies.}
    \label{grap:DopplerGeneralization}
\end{figure}

Fig.~\ref{grap:DopplerGeneralization} analyzes the generalization of the model with respect to the variation of Doppler spreads. \ac{MSE} values are compared for two training strategies: 1) \textit{model-switch training}, where a separate model is trained for each Doppler frequency, and 2) \textit{generalized training}, where a single model is trained on a dataset containing a mixture of Doppler conditions ranging from 1 to 50~Hz. The \ac{GRU}-based predictor demonstrates superior generalization in this context, achieving consistent \ac{MSE} performance across all Doppler scenarios using a unified model. In contrast, the Wiener filter shows degraded performance under generalized training, indicating a stronger dependency on Doppler-specific tuning.

\begin{figure}[htbp]
    \centering
    \includegraphics[width=0.9\linewidth]{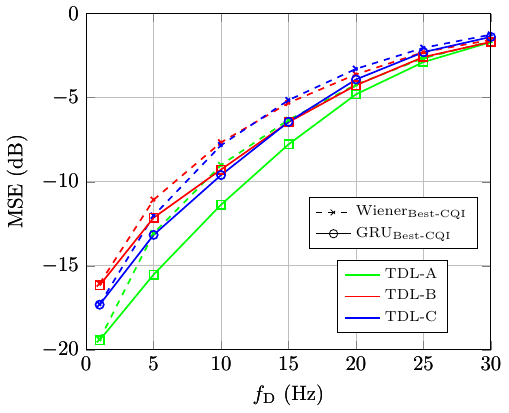}
    \caption{Comparison of MSE prediction error (in dB) for the Wiener and GRU models trained on the TDL-A channel and evaluated on different NLOS channel profiles: TDL-A, TDL-B, and TDL-C. }
    \label{grap:DelayProfileGeneralizationNLOS}
\end{figure}

Fig.~\ref{grap:DelayProfileGeneralizationNLOS} illustrates the generalization performance of the prediction models when evaluated on different channel profiles. Both the Wiener and \ac{GRU} models were trained on the TDL-A channel model and tested under three NLOS scenarios: TDL-A, TDL-B, and TDL-C. \ac{MSE}  is reported as a function of Doppler frequency \( f_D \), for the best-CQI case.

The results show that both predictors experience a performance degradation as \( f_D \) increases, consistent with the greater temporal variability of the channel. However, the \ac{GRU}-based model outperforms the Wiener predictor across most Doppler frequencies and channel profiles, with the largest gains observed in the low-to-moderate Doppler range (\( 5~\text{Hz} \leq f_D < 15~\text{Hz} \)). At very low Doppler values (e.g., \( f_D = 1~\text{Hz} \)), both predictors perform equally well.

Interestingly, although both models were trained on the TDL-A profile, they generalize relatively well to TDL-B and TDL-C. The gap \ac{MSE} between the profiles remains within 1--2~dB for each method, indicating that both frameworks are robust to moderate mismatches in the propagation environment. Among the three profiles, the TDL-A case produces the lowest \ac{MSE}, followed by TDL-C and then TDL-B, suggesting that the predictor trained on TDL-A generalizes effectively to propagation conditions with similar delay spreads.

These findings highlight the practical viability of using pre-trained models in deployment scenarios with varying NLOS characteristics, and further confirm the enhanced robustness of \ac{GRU}-based architectures for CSI prediction.

\begin{figure}[htbp]
    \centering
    \includegraphics[width=0.9\linewidth]{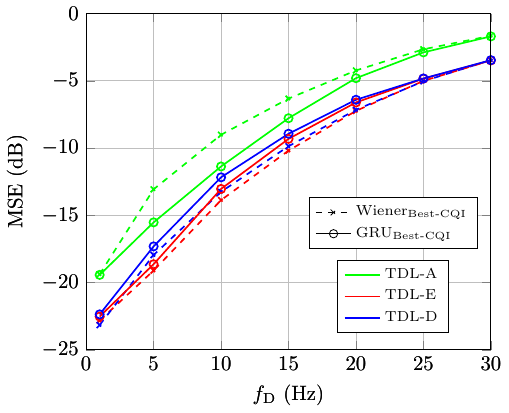}
    \caption{Comparison of MSE prediction error (in dB) for the Wiener and GRU models trained on the TDL-A channel and evaluated on different LOS channel profiles: TDL-A, TDL-E, and TDL-D.}
    \label{grap:DelayProfileGeneralizationLOS}
\end{figure}

Fig.~\ref{grap:DelayProfileGeneralizationLOS} evaluates the generalization performance of the prediction models under different LOS channel conditions. Both the Wiener and \ac{GRU} models were trained using the TDL-A profile and tested on three LOS channel models: TDL-A, TDL-E, and TDL-D. The \ac{MSE} is plotted as a function of Doppler frequency \( f_D \), and the prediction strategy corresponds to the best CQI.

Among the channel profiles evaluated, TDL-D and TDL-E achieve the lowest \ac{MSE} values, especially at moderate to high Doppler frequencies. This behavior can be attributed to their dominant line-of-sight components, which result in more predictable temporal channel variations. Surprisingly, the TDL-A profile leads to the highest prediction error, even though the models were trained on TDL-A. Its propagation characteristics appear less favorable for prediction compared to more structured profiles like TDL-D or TDL-E.

Overall, this figure demonstrates that both predictors remain robust under LOS channel conditions, with the Wiener-based model outperforming the \ac{GRU}. It also illustrates that models trained on general channel profiles, such as TDL-A, can generalize effectively to more deterministic environments.

\subsubsection{CSI Acquisition in FDD}
In \ac{FDD} systems, downlink \ac{CSI} must be explicitly estimated at the \ac{UE} and reported back to the base station, typically in the form of a single quantized \ac{CQI} value. In this subsection, we evaluate the performance of the proposed predictors in the FDD context. The primary goal is to determine the optimal future time instant, or prediction horizon \( \breve{\tau} \), at which the \ac{CQI} should be predicted to maximize system throughput.

\begin{figure}[htbp]
	\centering
	\includegraphics[width=0.9\linewidth]{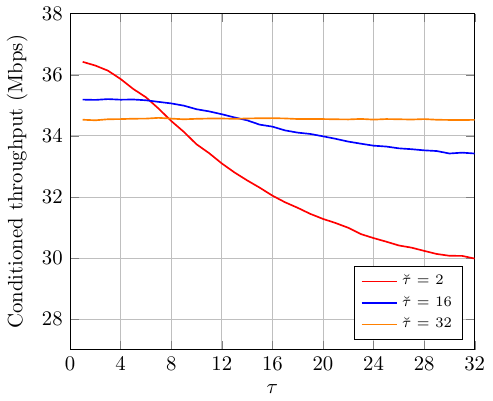}
	\caption{Instantaneous throughput per slot for the GRU-based predictor in FDD mode, shown for different prediction horizons \( \breve{\tau} \).}
	\label{fig:FDDInstantaneousThroughput}
\end{figure}

Our analysis reveals a counter-intuitive yet critical finding: the maximum average throughput is achieved not with a short-term prediction, but with a horizon \( \breve{\tau} > T_{\text{CSI}}/2 \). This behavior stems from a fundamental trade-off between immediate prediction accuracy and robustness against channel aging over the full transmission interval. This effect is clearly illustrated by the conditioned throughput shown in Fig.~\ref{fig:FDDInstantaneousThroughput}. For a short horizon (\( \breve{\tau} = 2 \)), the throughput starts high but decays rapidly as the predicted \ac{CQI} becomes outdated. Conversely, for longer horizons (\( \breve{\tau} = 16 \) or \( 32 \)), the initial throughput is lower, but it remains remarkably stable across the entire reporting window. This stability prevents severe performance degradation in the later slots, ultimately resulting in a higher overall average throughput.

\begin{figure}[htbp]
	\centering
	\includegraphics[width=0.9\linewidth]{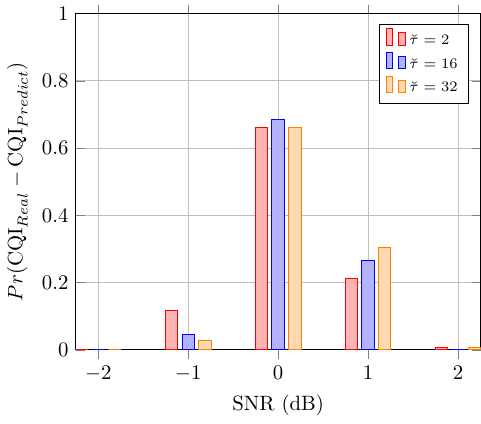}
	\caption{Probability distribution of the \ac{CQI} prediction error (defined as \( \text{CQI}_{\text{Real}} - \text{CQI}_{\text{Predicted}} \)) for different prediction horizons.}
	\label{fig:FDDCQISelection}
\end{figure}

The underlying reason for this stability is revealed in Fig.~\ref{fig:FDDCQISelection}, which shows the probability distribution of the \ac{CQI} prediction error. For longer prediction horizons, the model learns to be conservative. Observe that for \( \breve{\tau} = 16 \) and \( 32 \), the probability of making a perfect prediction (error = 0) decreases, but the probability of underestimating the \ac{CQI} by one level (a conservative error of +1) increases significantly. Crucially, the probability of overestimating the \ac{CQI} (an aggressive error of -1) remains very low. This conservative bias leads to the selection of a slightly more robust, lower-order \ac{MCS}. While this choice sub-optimally utilizes the channel in the initial slots (hence the lower initial throughput in Fig.~\ref{fig:FDDInstantaneousThroughput}), it prevents link failures in the later slots, ensuring consistent performance throughout the transmission interval.

\section{Discussion and Conclusion}
\label{sec:Conclusions}

This paper has presented and evaluated two \ac{CSI} prediction frameworks for \ac{TDD} and \ac{FDD} systems, operating in the effective \ac{SINR} domain to maintain compatibility with standard \ac{EESM}-based link adaptation while reducing computational complexity. A comparative analysis between a classical Wiener filter and a learning-based \ac{GRU} predictor was conducted under diverse channel conditions. The key findings and implications of this study are summarized as follows:

\begin{itemize}
    \item \textbf{Domain efficiency:} Operating in the effective \ac{SINR} domain significantly reduces dimensionality compared to predicting the full channel matrix, enabling efficient real-time implementation without compromising compatibility with 5G standards.
    
    \item \textbf{Optimal complexity:} An input sequence length of just four samples was found to be sufficient for both predictors. Increasing model depth or input history yields diminishing returns, particularly at high Doppler shifts.
    
    \item \textbf{Generalization vs. complexity:} The \ac{GRU}-based predictor outperforms the Wiener filter in \ac{MSE} and exhibits superior generalization capabilities across different channel profiles (e.g., training on TDL-A and testing on TDL-B/C). However, this comes at a higher computational cost. Surprisingly, the Wiener filter remains a competitive, ultra-low-complexity alternative, especially in very low or very high mobility regimes where complex learning offers limited gains.
    
    \item \textbf{FDD prediction horizon:} A counter-intuitive result was observed in \ac{FDD} systems: predicting a distant horizon (e.g., \( \breve{\tau} > T_{\text{CSI}}/2 \)) yields a higher average throughput than short-term prediction. This is due to the predictor learning a conservative bias that stabilizes performance over the entire reporting interval, preventing link failures caused by channel aging.
    
    \item \textbf{Deployment strategy:} The trade-off analysis suggests a split deployment strategy. The computationally intensive but robust \ac{GRU} approach is well-suited for \ac{TDD} \acp{BS}, where processing power is available. Conversely, the lightweight Wiener filter is the preferable choice for \ac{FDD} \acp{UE}, offering reasonable performance with minimal impact on device power consumption.
\end{itemize}

Overall, these findings provide practical guidelines for selecting \ac{CSI} prediction strategies based on specific system constraints, supporting more efficient and adaptive link management in future wireless networks.

\bibliographystyle{IEEEtran}
\bibliography{references.bib}

\begin{thebibliography}{10}
\providecommand{\url}[1]{#1}
\csname url@samestyle\endcsname
\providecommand{\newblock}{\relax}
\providecommand{\bibinfo}[2]{#2}
\providecommand{\BIBentrySTDinterwordspacing}{\spaceskip=0pt\relax}
\providecommand{\BIBentryALTinterwordstretchfactor}{4}
\providecommand{\BIBentryALTinterwordspacing}{\spaceskip=\fontdimen2\font plus
\BIBentryALTinterwordstretchfactor\fontdimen3\font minus \fontdimen4\font\relax}
\providecommand{\BIBforeignlanguage}[2]{{%
\expandafter\ifx\csname l@#1\endcsname\relax
\typeout{** WARNING: IEEEtran.bst: No hyphenation pattern has been}%
\typeout{** loaded for the language `#1'. Using the pattern for}%
\typeout{** the default language instead.}%
\else
\language=\csname l@#1\endcsname
\fi
#2}}
\providecommand{\BIBdecl}{\relax}
\BIBdecl

\bibitem{Vega2021}
F.~J. Martín-Vega, J.~C. Ruiz-Sicilia, M.~C. Aguayo, and G.~Gómez, ``Emerging tools for link adaptation on {5G NR} and beyond: Challenges and opportunities,'' \emph{IEEE Access}, vol.~9, pp. 126\,976--126\,987, 2021.

\bibitem{Goldsmith2005}
A.~Goldsmith, \emph{Wireless Communications}.\hskip 1em plus 0.5em minus 0.4em\relax Cambridge university press, 2005.

\bibitem{Yue2020}
Y.~Wang, W.~Liu, and L.~Fang, ``Adaptive modulation and coding technology in {5G} system,'' in \emph{Proc. International Wireless Communications and Mobile Computing (IWCMC)}, 2020, pp. 159--164.

\bibitem{38.214}
3GPP, \emph{{Technical Specification (TS); Physical layer procedures for data}}, {3rd Generation Partnership Project (3GPP)} TS {38.214}, Rev. 17.12.0, January 2025.

\bibitem{Papazafeiropoulos2017}
A.~K. Papazafeiropoulos, ``Impact of general channel aging conditions on the downlink performance of {Massive MIMO},'' \emph{IEEE Transactions on Vehicular Technology}, vol.~66, no.~2, pp. 1428--1442, 2017.

\bibitem{Truong2013}
K.~T. Truong and R.~W. Heath, ``Effects of channel aging in massive {MIMO} systems,'' \emph{Journal of Communications and Networks}, vol.~15, no.~4, pp. 338--351, 2013.

\bibitem{Jiang2017}
C.~Jiang and et~al, ``Machine learning paradigms for next-generation wireless networks,'' \emph{IEEE Wireless Communications}, vol.~24, no.~2, pp. 98--105, 2017.

\bibitem{Liu1995}
Y.~Liu and S.~Blostein, ``Identification of frequency non-selective fading channels using decision feedback and adaptive linear prediction,'' \emph{IEEE Transactions on Communications}, vol.~43, no. 2/3/4, pp. 1484--1492, 1995.

\bibitem{Jiang2019}
W.~Jiang and H.~D. Schotten, ``A comparison of wireless channel predictors: Artificial intelligence versus {Kalman} filter,'' in \emph{Proc. IEEE International Conference on Communications (ICC)}, 2019, pp. 1--6.

\bibitem{Kim2021}
H.~Kim and et~al, ``Massive {MIMO} channel prediction: Kalman filtering vs. machine learning,'' \emph{IEEE Transactions on Communications}, vol.~69, no.~1, pp. 518--528, 2021.

\bibitem{blanquez2016eolla}
F.~Blanquez-Casado and et~al, ``{eOLLA: an enhanced outer loop link adaptation for cellular networks},'' \emph{EURASIP Journal on Wireless Communications and Networking}, vol. 2016, no.~1, p.~20, 2016.

\bibitem{Ye2018}
H.~Ye, G.~Y. Li, and B.-H. Juang, ``Power of deep learning for channel estimation and signal detection in {OFDM} systems,'' \emph{IEEE Wireless Communications Letters}, vol.~7, no.~1, pp. 114--117, 2018.

\bibitem{Liao2019}
Y.~Liao and et~al, ``{CSI} feedback based on deep learning for {Massive MIMO} systems,'' \emph{IEEE Access}, vol.~7, pp. 86\,810--86\,820, 2019.

\bibitem{Saad2020}
W.~Saad, M.~Bennis, and M.~Chen, ``A vision of {6G} wireless systems: Applications, trends, technologies, and open research problems,'' \emph{IEEE Network}, vol.~34, no.~3, pp. 134--142, 2020.

\bibitem{2022Hong}
\BIBentryALTinterwordspacing
S.~Hong and et~al, ``Machine learning-based adaptive {CSI} feedback interval,'' \emph{ICT Express}, vol.~8, no.~4, pp. 544--548, 2022. [Online]. Available: \url{https://www.sciencedirect.com/science/article/pii/S2405959521001545}
\BIBentrySTDinterwordspacing

\bibitem{li2019ea}
Y.~Li and et~al, ``{EA-LSTM}: Evolutionary attention-based {LSTM} for time series prediction,'' \emph{Knowledge-Based Systems}, vol. 181, p. 104785, 2019.

\bibitem{Kadambar2023Deep}
S.~Kadambar and et~al, ``Deep learning based joint {CSI} compression and prediction for beyond-{5G} systems,'' in \emph{Proc. IEEE Global Communications Conference}, 2023, pp. 4792--4797.

\bibitem{2023Gao}
J.~Gao and et~al, ``Fast time-varying wireless channel prediction based on deep learning,'' in \emph{Proc. 9th International Conference on Computer and Communications (ICCC)}, 2023, pp. 940--945.

\bibitem{2021Yuan}
Z.~Yuan, K.~Niu, and C.~Dong, ``Channel prediction and {PMI/RI} selection in {MIMO-OFDM} systems based on deep learning,'' in \emph{Proc. IEEE 32nd Annual International Symposium on Personal, Indoor and Mobile Radio Communications (PIMRC)}, 2021, pp. 598--603.

\bibitem{Jian2022}
H.~Jiang, M.~Cui, D.~W.~K. Ng, and L.~Dai, ``Accurate channel prediction based on transformer: Making mobility negligible,'' \emph{IEEE Journal on Selected Areas in Communications}, vol.~40, no.~9, pp. 2717--2732, 2022.

\bibitem{Vaca2024}
\BIBentryALTinterwordspacing
C.~J. Vaca-Rubio and et~al, ``Kolmogorov-arnold networks (kans) for time series analysis,'' 2024. [Online]. Available: \url{https://arxiv.org/abs/2405.08790}
\BIBentrySTDinterwordspacing

\bibitem{Kadambar2023Smart}
S.~Kadambar and et~al, ``Smart-{CSI}: Deep learning based low complexity {CSI} prediction for beyond-{5G} systems,'' in \emph{Proc. IEEE 98th Vehicular Technology Conference (VTC2023-Fall)}, 2023, pp. 1--5.

\bibitem{Qin2022}
Z.~Qin and et~al, ``A partial reciprocity-based channel prediction framework for {FDD Massive MIMO} with high mobility,'' \emph{IEEE Transactions on Wireless Communications}, vol.~21, no.~11, pp. 9638--9652, 2022.

\bibitem{38.901}
3GPP, \emph{{Technical Report (TR); Study on channel model for frequencies from 0.5 to 100 GHz}}, {3rd Generation Partnership Project (3GPP)} TR {38.901}, Rev. 17.0.0, April 2022.

\bibitem{2010Tobias}
T.~L. Jensen, S.~Kant, J.~Wehinger, and B.~H. Fleury, ``Fast link adaptation for mimo ofdm,'' \emph{IEEE Transactions on Vehicular Technology}, vol.~59, no.~8, pp. 3766--3778, 2010.

\bibitem{Lagen2020}
S.~Lagen and et~al, ``New radio physical layer abstraction for system-level simulations of {5G} networks,'' in \emph{Proc. IEEE International Conference on Communications (ICC)}, 2020, pp. 1--7.

\bibitem{kay1993fundamentals}
S.~M. Kay, \emph{Fundamentals of statistical signal processing: estimation theory}.\hskip 1em plus 0.5em minus 0.4em\relax Prentice-Hall, Inc., 1993.

\bibitem{Brockwell1987}
P.~J. Brockwell and R.~A. Davis, \emph{{Time series: Theory and Methods}}.\hskip 1em plus 0.5em minus 0.4em\relax Springer-Verlag, 1987.

\bibitem{Mattu2022}
S.~R. Mattu and et~al, ``{Deep Channel Prediction: A DNN Framework for Receiver Design in Time-Varying Fading Channels},'' \emph{IEEE Transactions on Vehicular Technology}, vol.~71, no.~6, pp. 6439--6453, 2022.

\bibitem{Hochreiter1997LSTM}
S.~Hochreiter and J.~Schmidhuber, ``{Long Short-Term Memory},'' \emph{Neural Computation}, vol.~9, no.~8, pp. 1735--1780, 1997.

\bibitem{chung2014empiricalevaluationgatedrecurrent}
\BIBentryALTinterwordspacing
J.~Chung and et~al, ``Empirical evaluation of gated recurrent neural networks on sequence modeling,'' 2014. [Online]. Available: \url{https://arxiv.org/abs/1412.3555}
\BIBentrySTDinterwordspacing

\bibitem{3gpp_ran112}
\BIBentryALTinterwordspacing
F.~J. Martín-Vega, M.~C. Aguayo-Torres, G.~Gómez, J.~Campos, and J.~Torrecilla, ``R1-2300071, ``{Further discussions of AI/ML for CSI feedback enhancement}'','' in \emph{3GPP TSG RAN WG1 Meeting \#112, Athens, Greece}, February 27th – March 3rd, 2023, pp. 1--14. [Online]. Available: \url{https://www.3gpp.org/ftp/tsg_ran/WG1_RL1/TSGR1_112/Docs/R1-2300071.zip}
\BIBentrySTDinterwordspacing

\end{thebibliography}
\end{document}